\documentclass[pra,final,twocolumn,showpacs,showkeys]{revtex4}
\usepackage{amsmath}
\usepackage{graphicx}
\usepackage{amsfonts}
\usepackage{amssymb}
\usepackage{subfigure}

\newcommand{\subs}[1]
{ 
	\mbox{\scriptsize{#1}}
}

\begin{document}

\title{Entangled State Synthesis for Superconducting Resonators}

\author{Frederick W. Strauch$^1$} \email[Electronic address: ]{Frederick.W.Strauch@williams.edu}
\author{Douglas Onyango$^1$}
\author{Kurt Jacobs$^2$}
\author{Raymond W. Simmonds$^3$ }

\affiliation{$^1$Williams College, Williamstown, MA 01267, USA \\
$^2$Department of Physics, University of Massachusetts at Boston, 100 Morrissey Blvd., Boston, Massachusetts 02125, USA \\
$^3$National Institute of Standards and Technology, 325 Broadway, Boulder, Colorado 80305 USA}

\date{\today}

\begin{abstract}
We present a theoretical analysis of methods to synthesize entangled states of two superconducting resonators.  These methods use experimentally demonstrated interactions of resonators with artificial atoms, and offer efficient routes to generate nonclassical states.  We analyze physical implementations, energy level structure, and the effects of decoherence through detailed dynamical simulations.  

\end{abstract}
\maketitle
%%%%End of Front Matter%%%%%%%%%%%%%%%%%%%%%%%%%%%%%%%%%%%%%%%%%%%%%%%%%
%%%%%%%%%%%%%%%%%%%%%%%%%%%%%%%%%%%%%%%%%%%%%%%%%%%%%%%%%%%%%%%%%%%%%%%%

\section{Introduction}

The control of quantum systems and their applications have rejuvenated the study of natural and artificial atomic systems.  One of the most exciting such studies is the light-matter interaction, exemplified by cavity-QED and the Jaynes-Cummings Hamiltonian \cite{HarocheBook}.  In particular, the use of superconducting resonators, with harmonic oscillator modes of the electromagnetic field in the microwave domain, has gone through a remarkable transformation in the past decade.  

Early on, harmonic oscillator modes were incorporated in superconducting qubit circuits as way to couple them together, and thus perform joint operations.  The first such proposal~\cite{Shnirman97} used a dispersive coupling in which the qubit frequency, $\omega_q$, is much less than the resonator frequency, $\omega_r$, and this design was refined in~\cite{Makhlin99}.  Resonant coupling was first discussed in~\cite{Makhlin01}, and yields an architecture that is very similar to the ion-trap quantum computer~\cite{Cirac95}, in which the resonator is the analogue of the center-of-mass mode of the linear ion chain.  The qubit's quantum state can be transferred to the oscillator, which then interacts with other qubits, and is finally transferred back. Resonant coupling was later adapted to couple charge qubits~\cite{Plastina2003} and flux qubits~\cite{Smirnov2002} via an LC-oscillator, and charge qubits via a current-biased junction~\cite{Blais2003,Wei05}.  In parallel with these fundamental studies came the observation that these systems provide strong analogies with cavity quantum electrodynamics (QED)~\cite{Yang03, You03}.  This general approach has now come to be called circuit-QED~\cite{Blais04}, and has led to impressive experimental progress~\cite{Wallraff04, Chiorescu04, Xu2005,Johansson06, Sillanpaa2007, Schuster07, Majer07, Houck2008, DiCarlo2009}.  

The use of on-chip superconducting resonators to measure and couple qubits was then extended to include ideas for producing a quantum memory element and bus for transferring information, with either microwave \cite{Koch06, Sillanpaa2007} or even nanomechanical \cite{Cleland04,Geller05, Pritchett05} modes.  This has culminated in the ``von Neumann'' model of quantum logic by Mariantoni \textit{et al.} \cite{Mariantoni11}, consisting of a qubit-based processing unit, a resonator-based memory, and a resonator data bus.  However, while the nonlinearity of a qubit or other auxiliary is needed to access the individual states of the resonator, there remain alternatives to qubit-based logic.  

Quantum systems with $d>2$ logical levels are traditionally called {\it{qudits}}, and unitary gate synthesis \cite{Brennen05} and error correction \cite{Gottesman99} can be constructed using qudits.  The Fock states of a superconducting resonator can be addressed as a qudit \cite{Strauch11}, with the potential to out-perform qubit-based processing.  It is the purpose of this paper to study one of the simplest such tasks in quantum information processing, namely the generation of entanglement between two systems.  While there is essentially just one type of entangled state between two qubits, qudits can be entangled in any of the states
\begin{equation}
|\Psi \rangle = \sum_{n=0}^{N} c_n |n\rangle \otimes |N-n\rangle,
\end{equation}
with an entanglement up to $\log_2 (N+1)$ ebits \cite{Bennett96}.  We shall study the synthesis of the maximally entangled two-qudit state (with $c_n = 1/\sqrt{N+1}$) and the ``NOON'' state
\begin{equation}
|\Psi_{\subs{NOON}} \rangle = \frac{1}{\sqrt{2}} \left( |N\rangle \otimes |0\rangle + |0\rangle \otimes |N\rangle \right),
\end{equation}
a state of interest for quantum metrology \cite{Dowling08} and recently demonstrated \cite{Wang11} in a superconducting circuit.

The NOON state experiment followed the implementation~\cite{Hofheinz2009} of the quantum state synthesis algorithm for a single oscillator coupled to a qubit proposed by Law and Eberly~\cite{Law96}.  In this implementation, a superconducting phase qubit was capacitively coupled to a superconducting resonator.  By performing a sequence of ``shift pulses'' and Rabi pulses, the qubit was put into resonance with the resonator, thus transferring quanta between the two systems.  It was shown in~\cite{Law96} that a properly chosen sequence of such operations could generate an arbitrary state of the resonator. In the experiment of Hofheinz \textit{et al.}, several quantum states were prepared and analyzed by Wigner function tomography~\cite{Hofheinz2009}.  An extension of this method, using two three-level systems coupled to two resonators \cite{Merkel10}, was used to generate the NOON state with $N=3$ \cite{Wang11}.

An alternative approach was previously proposed by the authors \cite{Strauch10} to synthesize an arbitrary entangled state of two resonators.  Our algorithm was based on previous studies of Law-Eberly-like schemes to generate two-mode quantum states of the motional degrees of freedom of a trapped ion~\cite{Gardiner97, Kneer98}, suitably modified to use the interactions available in superconducting circuits.  In particular, the number-state-dependent transitions in the quasi-dispersive regime demonstrated by Johnson \textit{et al.} \cite{Johnson10} played a critical role in allowing selective manipulations of the Fock states of the two resonators.  

An important issue missing from previous analyses \cite{Strauch10,Merkel10} is the role of decoherence in these methods.  The coherence of the coupled system is influenced by decoherence in both the qubit and the resonator, and understanding these effects is an important theoretical question.  In this paper we provide a complete analysis of the state synthesis algorithm and its application to the entangled states mentioned above.  We will further discuss the experimental issues, including decoherence, that arise in the synthesis of NOON states, and compare the fidelity of the results of the state-synthesis algorithm \cite{Strauch10} with the NOON state sequence of \cite{Merkel10}.  The methods and results of this study provide a solid base for understanding more complex manipulations of superconducting resonators as qudits \cite{Strauch11}.

This paper is organized as follows.  In Section II the state-synthesis algorithm is explored in some detail, with special attention given to physical implementation.  In Section III, we provide a numerical simulation of NOON-state synthesis in the strong-coupling regime, illustrating some of the subtleties of implementation.  In Section IV, we present results of NOON-state synthesis in the presence of decoherence.  Finally, we conclude in Section V, while details of decoherence calculations are saved for an Appendix.  

\section{State Synthesis Algorithm}
\subsection{Physical Model}
Our algorithm is designed with the following model Hamiltonian
\begin{equation}
\mathcal{H} = \mathcal{H}_{q}(t) + \mathcal{H}_{r} +  \mathcal{H}_{a} + \mathcal{H}_{b}
\end{equation}
with
\begin{eqnarray}
\mathcal{H}_q(t) &=& \hbar \omega_q(t) \sigma_+ \sigma_-  +  \frac{1}{2} \hbar \left[ \Omega(t) \sigma_+ + \Omega^*(t) \sigma_- \right],  \\
\mathcal{H}_{r} &=& \hbar \omega_a a^{\dagger} a + \hbar \omega_b b^{\dagger} b ,\\
\mathcal{H}_{a} &=& \hbar g_a \left( \sigma_+ a  + \sigma_- a^{\dagger} \right), \\
\mathcal{H}_{b} &=& \hbar g_b \left( \sigma_+ b  + \sigma_- b^{\dagger} \right),
\end{eqnarray} 
where $\sigma_+ = |1\rangle \langle 0|, a^{\dagger}$, and $b^{\dagger}$ are the creation operators for the qubit, resonator $a$, and resonator $b$, respectively.  Our goal is to control an arbitrary state of the resonator, implementing the transformation
\begin{equation}
U |0,0,0\rangle = \sum_{n_a,n_b} c_{n_a,n_b} |0,n_a,n_b\rangle,
\end{equation}
where $|q,n_a,n_b\rangle = |q\rangle \otimes |n_a\rangle \otimes |n_b\rangle$ is the state in which the qubit is in state $q=0$ or $1$, resonator $a$ is in Fock state $n_a$, and resonator $b$ is in Fock state $n_b$.   All controls are implemented by manipulating the qubit by shifts of the qubit frequency $\omega_q(t)$ or application of resonant Rabi pulses of the form $\Omega(t) = \Omega_0 e^{i \omega_d t}$.   We will shift the qubit frequency from the dispersive regime, with $\omega_a < \omega_q < \omega_b$ (generalizations will be discussed later), to the resonant regimes $\omega_q = \omega_a$ or $\omega_q = \omega_b$.  In this section we will present the theoretical background and motivation for the state-synthesis algorithm.  An explicit application will be presented in the following section.

In the dispersive regime $g_a \ll |\omega_q-\omega_a|$, $g_b \ll |\omega_q-\omega_b|$, the states $|q,n_a,n_b\rangle$ are approximate eigenstates.  The Hamiltonian couples each such state to two others (i.e. $|0,n_a,n_b\rangle \to |1,n_a-1,n_b\rangle, |1,n_a,n_b-1\rangle$ and $|1,n_a,n_b\rangle \to |0,n_a+1,n_b\rangle, |0,n_a,n_b+1\rangle$).  Perturbation theory yields the following approximation for the energies $E_{q,n_a, n_b}$:
\begin{eqnarray}
E_{0,n_a,n_b}/\hbar &\approx& \omega_a n_a + \omega_b n_b + \frac{g_a^2}{\omega_a - \omega_q} n_a  \nonumber \\
& & + \quad \frac{g_b^2}{\omega_b - \omega_q} n_b, \\
E_{1,n_a,n_b}/\hbar &\approx& \omega_{q} + \omega_a n_a+  \omega_b n_b +  \frac{g_a^2}{\omega_q - \omega_a} (n_a+1) \nonumber \\
 & & + \quad \frac{g_b^2}{\omega_q - \omega_b} (n_b+1) .
\end{eqnarray}
The drive frequency needed for the transition $|0,n_a,n_b\rangle \to |1,n_a,n_b\rangle$ is thus given by
\begin{eqnarray}
\omega_d &=& (E_{1,n_a,n_b} - E_{0,n_a,n_b})/\hbar \nonumber \\
& \approx& \omega_q + (2n_a+1) \frac{g_a^2}{\omega_q-\omega_a} + (2n_b+1) \frac{g_b^2}{\omega_q-\omega_b} \nonumber \\
\label{drivefreq}
\end{eqnarray}

This Stark shift can be used to selectively address individual states $(n_a,n_b)$, conveniently illustrated by a Fock-state diagram as in Fig. \ref{Fock_Diagram}.  Here we have set $g_a^2/(\omega_q - \omega_a) = g_b^2/(\omega_b - \omega_q) = \Delta \omega/2$ \footnote{Reference \cite{Strauch10} had a factor of two error here.}, allowing us to simplify the drive frequencies to
\begin{equation}
\omega_n = \omega_q + n \Delta \omega.
\end{equation} 
For this drive frequency, only those states for which $n_a - n_b = n$ is constant will undergo the transition $|0,n_a,n_b\rangle \to |1,n_a,n_b\rangle$.  Note that this assumes a two-level system for which $\omega_a < \omega_q < \omega_b$.  As discussed below, this condition is not strictly necessary for the state-synthesis algorithm.  The key requirement for an efficient state synthesis approach is to have an interaction such that a given Fock-state $(n_a,n_b)$ can be addressed independently of the neighboring states $(n_a \pm 1, n_b \pm 1)$ \cite{Kneer98}.

We label the $\omega_d = \omega_n$ transition $R_{01}^{(n)}$.   Letting $|\Psi\rangle$ be given by
\begin{equation}
|\Psi\rangle = \sum_{q,n_a,n_b} c_{q,n_a,n_b} |q,n_a,n_b\rangle,
\end{equation}
the transition has the following effect:
\begin{equation}
R_{01}^{(n)} |\Psi\rangle = \sum_{q,n_a,n_b} r_{q,n_a,n_b} |q,n_a,n_b\rangle
\end{equation}
where for $n_a - n_b = n$:
\begin{equation}
\begin{array}{lcl}
r_{0,n_a,n_b} &=& e^{i \alpha} \cos (\theta/2)  c_{0,n_a,n_b} - i e^{i \beta} \sin (\theta/2) c_{1,n_a,n_b} \\
r_{1,n_a,n_b} &=& e^{-i \alpha} \cos (\theta/2)  c_{1,n_a,n_b} - i e^{-i \beta} \sin (\theta/2) c_{0,n_a,n_b},
\end{array}
\end{equation}
where $\theta = \Omega_0 t$ and $r_{q,n_a,n_b} = e^{i \phi_q} c_{q,n_a,n_b}$ for $n_a - n_b \ne n$.  Here we have introduced the phases $(\alpha, \beta, \phi_q)$ in the transition.  As discussed below, these phases are needed to properly zero out amplitudes during the algorithm.  This phase control can be implemented by applying short shift pulses \cite{Hofheinz2009,DiCarlo2010,Mariantoni11}, by an amount $\Delta \omega_q$ for a duration $\tau$, providing a controllable phase shift $\Delta \omega_q \tau$.   The microwave pulse also has a controllable phase $\delta$, in terms of which $\alpha = \phi_0 = -\phi_1 = \Delta \omega_q \tau/2$  and $\beta = \delta + \Delta \omega_q \tau/2$.  However, for the states discussed here, such phase control is not necessary, and we will assume $\alpha = \beta = \phi_q = 0$.

By shifting the qubit's frequency into resonance with resonator $a$ ($\omega_q = \omega_a$) or resonator $b$ ($\omega_q = \omega_b$), the interaction terms can be made dominant.  Letting $A = e^{-i \mathcal{H}_{a} t/\hbar}$, we let
 \begin{equation}
A |\Psi\rangle = \sum_{q,n_a,n_b} a_{q,n_a,n_b} |q,n_a,n_b\rangle.
\end{equation}
The operation $A$ performs the mapping 
\begin{equation}
\begin{array}{lcl}
a_{0,n_a,n_b} &=&  \cos \theta_{n_a} c_{0,n_a,n_b} - i \sin \theta_{n_a} c_{1,n_a-1,n_b} \\
a_{1,n_a-1,n_b} &=& \cos \theta_{n_a} c_{1,n_a-1,n_b} - i \sin \theta_{n_a} c_{0,n_a,n_b},
\end{array}
\end{equation}
where $\theta_{n_a} = \sqrt{n_a} g_a t$. Similarly, letting $B = e^{-i \mathcal{H}_{b} t/\hbar}$, we find that
\begin{equation}
B |\Psi\rangle = \sum_{q,n_a,n_b} b_{q,n_a,n_b} |q,n_a,n_b\rangle
\end{equation}
performs the mapping
\begin{equation}
\begin{array}{lcl}
b_{0,n_a,n_b} &=& \cos \theta_{n_b} c_{0,n_a,n_b} - i \sin \theta_{n_b} c_{1,n_a,n_b-1} \\
b_{1,n_a,n_b-1} &=& \cos \theta_{n_b} c_{1,n_a,n_b-1} - i \sin \theta_{n_b} c_{0,n_a,n_b},
\end{array} 
\end{equation}
where $\theta_{n_b} = \sqrt{n_b} g_b t$.  Note that these three interactions provide all the two-level rotations needed between sets of qubit-resonator states.

\subsection{Algorithm}
\begin{figure}[t]
\begin{center}
\includegraphics[width=3in]{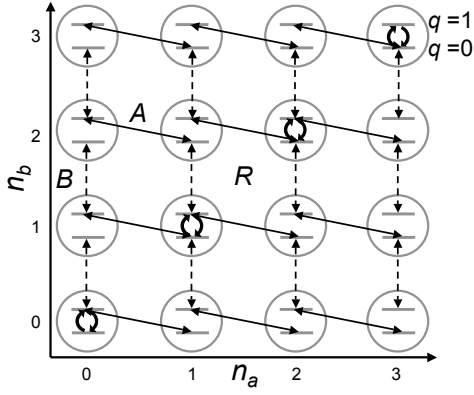}
\caption{Schematic set of operations to generate an arbitrary state of two coupled resonators.  In this Fock state diagram, each node corresponds to a specific set of photon numbers $(n_a,n_b)$, and contains two levels for the qubit.  Interactions lead to couplings between these states, indicated by the arrows between the states.  There are three key interactions used in this scheme.  $A$ transfers quanta between the qubit and resonator $a$ by a resonant Jaynes-Cummings-type interaction (solid lines).  Similary, $B$ transfers quanta between the qubit and resonator $b$ (dashed lines).  Finally $R$ (curved arrows) rotates the qubit for states with $n_a - n_b = n$ (here with $n=0$, see text).  }
\label{Fock_Diagram}
\end{center}
\end{figure}

These three interactions, $R_{01}^{(n)}$, $A$, and $B$, are illustrated in Fig. \ref{Fock_Diagram}.  By manipulating the qubit frequencies and Rabi drives, these three interactions can be effectively turned on and off sequentially for varying durations of time, directing population around the Fock-state diagram.  By starting from $|0,0,0\rangle$, the target state
\begin{equation}
 |\Psi_{\subs{target}}\rangle = U |0,0,0\rangle = \sum_{n_a,n_b} c_{n_a,n_b} |0,n_a,n_b\rangle,
 \end{equation}
 can be reached in a number of steps less than or equal to  $2 N_{\subs{max}} (N_{\subs{max}} + 1)$, where the amplitudes $c_{n_a,n_b}$ are nonzero for $n_a, n_b \le N_{\subs{max}} $.  

The algorithm is given by the following sequence of operations 
\begin{equation}
 U = \left(\prod_{j=1}^{N_b} U_{b,j} \right) U_{a}
 \end{equation}
 with
\begin{equation}
U_{a} =  \prod_{j=1}^{N_a} A_j R_{a,j}, \  \mbox{and} \ \ U_{b,j} =  \prod_{k=0}^{N_a} B_{jk} R_{b,jk}.
\end{equation}
The parameters ($t_{a,j}$, $t_{b,jk}$, $R_{a,j}$, and $R_{b,jk}$) in this operation depend on the original amplitudes $c_{n_a,n_b}$ and are found by solving the inverse evolution: 
\begin{equation}
U^{\dagger} |\Psi\rangle = U_{a}^{\dagger} \prod_{j=N_b}^1 U_{b,j}^{\dagger} |\Psi_{\subs{target}}\rangle = |0,0,0\rangle.
\end{equation} 
As in the state synthesis algorithm of Law and Eberly \cite{Law96,Kneer98}, each operation is a two-state rotation (specified above).  The (inverse) algorithm can then be visualized as a sequence of transitions on the Fock-state diagram, each step chosen to zero out a particular element $|q,n_a,n_b\rangle$.   The solution presented here moves amplitudes from right to left column-by-column, from top to bottom row-by-row, until finally all of the amplitude is located at $|0,0,0\rangle$.  Inverting this sequence produces the desired transformation $U |0,0,0\rangle \to |\Psi_{\subs{target}}\rangle$.  

In more detail, we have
\begin{equation}
U^{\dagger} = U_{a}^{\dagger} U_{b,1}^{\dagger} U_{b,2}^{\dagger} \cdots U_{b,N_b}^{\dagger},
\end{equation}
where 
\begin{equation}
U_{a}^{\dagger} = R_{a,1}^{\dagger} A_1^{\dagger} R_{a,2}^{\dagger} A_2^{\dagger} \cdots R_{a,N_a}^{\dagger} A_{N_a}^{\dagger}
\end{equation}
and
\begin{equation}
U_{b,j}^{\dagger} = R_{b,j0}^{\dagger} B_{j0}^{\dagger}  R_{b,j1}^{\dagger} B_{j1}^{\dagger} \cdots R_{j N_a}^{\dagger} B_{j N_a}^{\dagger}.
\end{equation}

\begin{figure}[t]
\begin{center}
\includegraphics[width=3in]{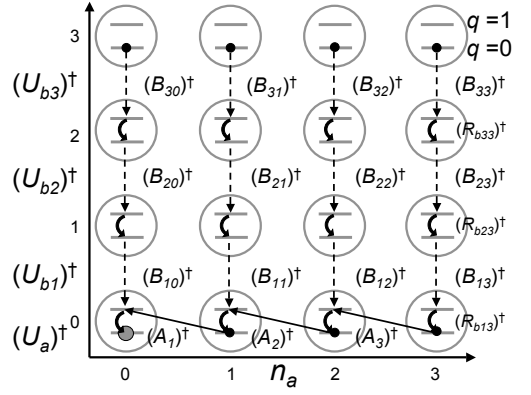}
\caption{Schematic set of steps in the (inverse of the) state-synthesis algorithm.  Not all of the Rabi rotations are shown.  }
\label{algorithm}
\end{center}
\end{figure}

The form of $U_{b,j}^{\dagger}$ is chosen to transfer all of the amplitude of the states in row $n_b = j$ to $n_b = j-1$.  For each $k$, $B_{b,jk}^{\dagger}$ transfers the amplitude from $|0,k,j\rangle$ to $|1,k,j-1\rangle$, after which $R_{b,jk}^{\dagger}$ rotates the state to $|0,k,j-1\rangle$.  The sequence of operations in $U_{b,j}^{\dagger}$ can thus be visualized as clearing out row $j$ in the Fock-state diagram, stepping from $k=N_a$ down to $k=0$.  This procedure is then repeated in $U^{\dagger}$ until all of the amplitude is in row $j=0$ and thus resonator $b$ is in the ground state ($n_b = 0$). The final sequence of operations in $U_a^{\dagger}$ is equivalent to the state-synthesis procedure of Law and Eberly \cite{Law96} for a qubit coupled to a single resonator (as modified in \cite{Hofheinz2009}).  This is illustrated in Fig. \ref{algorithm} for $N_a = N_b = 3$.
 
%%%%%%%%%%%%%%%%%%%%%%%%%%%%%%%%%%%%%%%%%
%%%%%%%%%%%%%%%%%%%%%%%%%%%%%%%%%%%%%%%%%

%%%%%%%%%%%%%%%%%%%%%%%%%%%%%%%%%%%%%%%%
%%%%%%%%%%%%%%%%%%%%%%%%%%%%%%%%%%%%%%%%%

%%%%%%%%%%%%%%%%%%%%%%%%%%%%%%%%%%%%%%%%%
\begin{figure}[t]
\begin{center}
\includegraphics[width=3in]{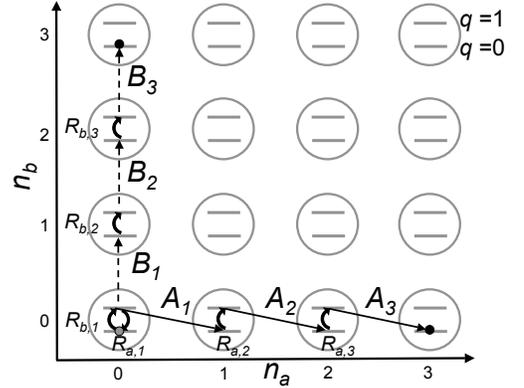}
\caption{Algorithm to generate the state $|\Psi \rangle = |3,0\rangle + |0,3\rangle$ of two coupled resonators.  The sequence of operations is detailed in Table 1.  The horizontal, vertical, and curved transitions are the interactions $A$, $B$, and $R$ (see text).}
\label{noonstatefig}
\end{center}
\end{figure}
%%%%%%%%%%%%%%%%%%%%%%%%%%%%%%%%%%%%%%%%%

An important issue not discussed in \cite{Strauch10} is the issue of relative phases of the various states.  In fact, since $\omega_a < \omega_q < \omega_b$, all of the states will evolve in time, so it is most convenient to specify the algorithm using an interaction picture with respect to the Hamiltonian $\mathcal{H}_q(0) + \mathcal{H}_r$, as above.   However, in the course of the algorithm it is necessary to zero out the amplitude of a particular state by rotating into another state.  For example, given a qubit state
\begin{equation}
|\psi\rangle = \cos \theta |0\rangle + e^{-i \phi} \sin \theta |1\rangle,
\end{equation}
the amplitude for state $|1\rangle$ can be removed by applying the rotation $e^{i \theta \sigma_y} e^{-i \phi \sigma_z/2}$ or $e^{i \theta \sigma_x} e^{-i (\phi - \pi/2) \sigma_z/2}$.  Both choices require shifting the relative phases between $|0\rangle$ and $|1\rangle$.  Fortunately, it is always the case that the two states coupled by $A$, $B$, and $R$ differ in the qubit excitation.  Thus, one can adjust the relative phases of the two relevant states by a short duration shift of the qubit frequency as described above. These small shifts can therefore be included between each of the various operations in $U$.    Specifying these phases is not necessary for the following examples.

The total number of operations in $U$ involves $N_a$ $A$ unitaries, $(N_a+1) N_b$ $B$ unitaries, and $N_a + (N_a + 1) N_b$ Rabi pulses $R$.   Correspondingly, the maximum total time for this algorithm is \cite{endnote54}
\begin{eqnarray}
T_{\subs{max}} &=& (N_a +N_b + N_a N_b)  \frac{\pi}{\Omega} + \sum_{j=1}^{N_a} \frac{\pi}{2 g_a \sqrt{j}} \nonumber \\
& & + (N_a+1) \sum_{j=1}^{N_b} \frac{\pi}{2 g_b \sqrt{j}} 
\label{tmax}
\end{eqnarray}
This assumes that all of the Rabi and shift pulses are $\pi$-pulses.  Of course,  specific instances of the algorithm can have times less than $T_{\subs{max}}$, as will be seen in the following examples.

\subsection{Specific Examples}

The simplest example is the construction of a NOON state
\begin{equation}
|\Psi_{\subs{NOON}}\rangle = \frac{1}{\sqrt{2}} |0\rangle \otimes \left( |N,0\rangle + |0,N\rangle \right).
\end{equation}
This state is particularly nice, as the solution for the inverse evolution admits the following simplification
\begin{equation}
U^{\dagger}_{b,j} = R_{b,j0}^{\dagger} B^{\dagger}_{j0}.
\end{equation}
The evolution times are chosen to move population from $|0,0,j\rangle \to |1,0,j-1\rangle \to |0,0,j-1\rangle$.  This is done for $j=N \to 1$, after which $U_a^{\dagger}$ moves population from $|0,N,0\rangle$ to $|0,0,0\rangle$.  Each step is a two-state rotation, and only trivial phases appear.  We can schematically write this as $U = (A R)^N (B R)^N$, where all of the rotations have $\theta = \pi$ except for the first $R$, which is a $\pi/2$-pulse.  The number of steps is $4N$, with a time \cite{endnote54}
\begin{eqnarray}
T_{\subs{NOON}} &=& \left( 2 N - \frac{1}{2} \right)  \frac{\pi}{\Omega} + \sum_{j=1}^{N} \frac{\pi}{2 g_a \sqrt{j}} \nonumber \\
& & + \sum_{j=1}^{N} \frac{\pi}{2 g_b \sqrt{j}} 
\label{tmax}
\end{eqnarray}
An explicit list of parameters for the twelve steps needed for $N=3$ is given in Table \ref{fockprogram1} 

\begin{table}
\caption{Procedure for $\Psi = |0, 3,0\rangle + |0,0,3\rangle$}. % title of Table 
\centering      % used for centering table 
\begin{tabular}{l l r}  % centered columns (4 columns) 
\hline\hline                        %inserts double horizontal lines 
Step & Parameters & Quantum State \\ [0.5ex] % inserts table 
%heading 
\hline                    % inserts single horizontal line 
$R_{a,1} $ & $\Omega t_{qa,1}= \pi/2, \omega_d = \omega_0$ & $|0,0,0\rangle - i |1,0,0\rangle$  \\ 
$A_{1} $ & $g_a t_{a,1}= \pi/2$ & $|0,0,0\rangle -  |0,1,0\rangle$  \\ 
$R_{a,2} $ & $\Omega t_{qa,2}= \pi , \omega_d = \omega_1$ & $|0,0,0\rangle + i |1,1,0\rangle$  \\ 
$A_{2} $ & $g_a t_{a,2}= \pi/(2\sqrt{2})$ & $|0,0,0\rangle +  |0,2,0\rangle$  \\ 
$R_{a,3} $ & $\Omega t_{qa,3}= \pi , \omega_d = \omega_2$ & $|0,0,0\rangle - i |1,2,0\rangle$  \\ 
$A_{3} $ & $g_a t_{a,3}= \pi/(2\sqrt{3})$ & $|0,0,0\rangle -  |0,3,0\rangle$  \\ 
$R_{b,1} $ & $\Omega t_{qb,1}= \pi, \omega_d = \omega_0$ & $-i |1,0,0\rangle - |0,3,0\rangle$  \\ 
$B_{1} $ & $g_b t_{b,1}= \pi/2$ &  $-|0,0,1\rangle -  |0,3,0\rangle$  \\ 
$R_{b,2} $ & $\Omega t_{qb,2}= \pi, \omega_d = \omega_{-1}$ & $i |1,0,1\rangle - |0,3,0\rangle$  \\ 
$B_{2} $ & $g_b t_{b,2}= \pi/(2\sqrt{2})$ & $|0,0,2\rangle -  |0,3,0\rangle$  \\ 
$R_{b,3} $ & $\Omega t_{qb,3}= \pi, \omega_d = \omega_{-2}$ & $-i |1,0,2\rangle - |0,3,0\rangle$  \\ 
$B_{3} $ & $g_b t_{b,3}= \pi/(2\sqrt{3})$ & $-|0,0,3\rangle -  |0,3,0\rangle$  \\ 
[1ex]       % [1ex] adds vertical space 
\hline     %inserts single line 
See \cite{endnote54}.
\end{tabular} 
\label{fockprogram1}  % is used to refer this table in the text 
\end{table} 

\begin{figure*}
\begin{center}
\includegraphics[width=7in]{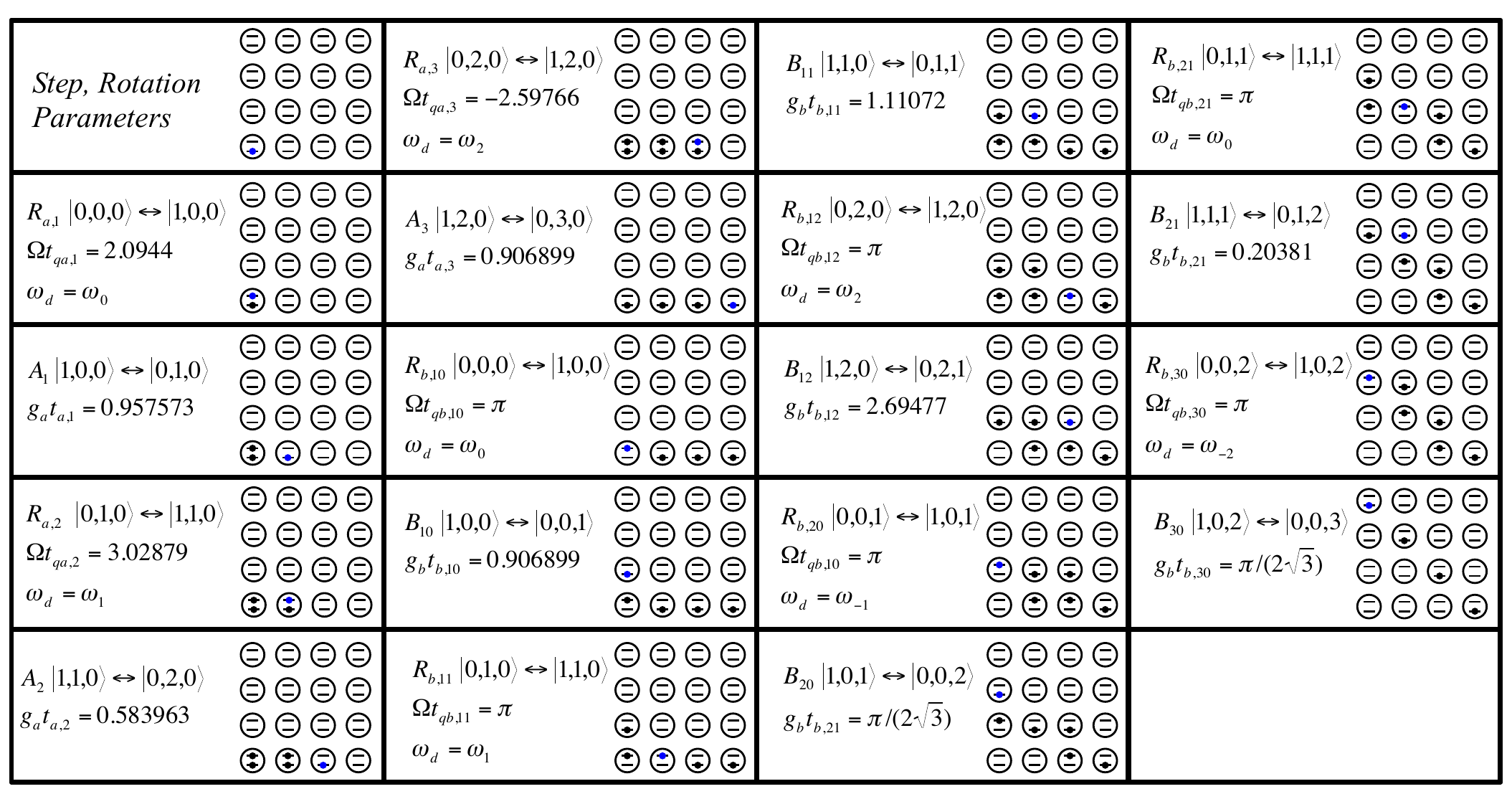}
\caption{Procedure for $\Psi = \sum_{n=0}^3 |0,n,3-n\rangle$.  Each block includes the step of the sequence, the parameter values, and the relevant transition being performed.  In addition, the state populations after the step are shown as a Fock-state diagram, with the relevant state highlighted in blue.  The parameters were found by numerically solving for the inverse evolution equation.   The forward sequence reads from top-to-bottom, left-to-right, while the inverse sequence is reversed.}
\label{fockprogram2}
\end{center}
\end{figure*}

A more complicated example is given by the maximally entangled state
\begin{equation}
|\Psi \rangle = \frac{1}{\sqrt{N+1}} \sum_{n=0}^{N} |n\rangle \otimes |N-n\rangle.
\end{equation}
While this only occupies a sparse region of the Fock-state diagram, the algorithm does not have the same simplifications or explicit solution as the NOON state.  Figure \ref{fockprogram2} shows a numerical solution for the eighteen steps of the control sequence for $U$ with $N=3$.  

%%%%%%%%%%%%%%%
%%%%%%%%%%%%%%%

\subsection{Superconducting Implementation}
\begin{figure}[b]
\begin{center}
\includegraphics[width=3in]{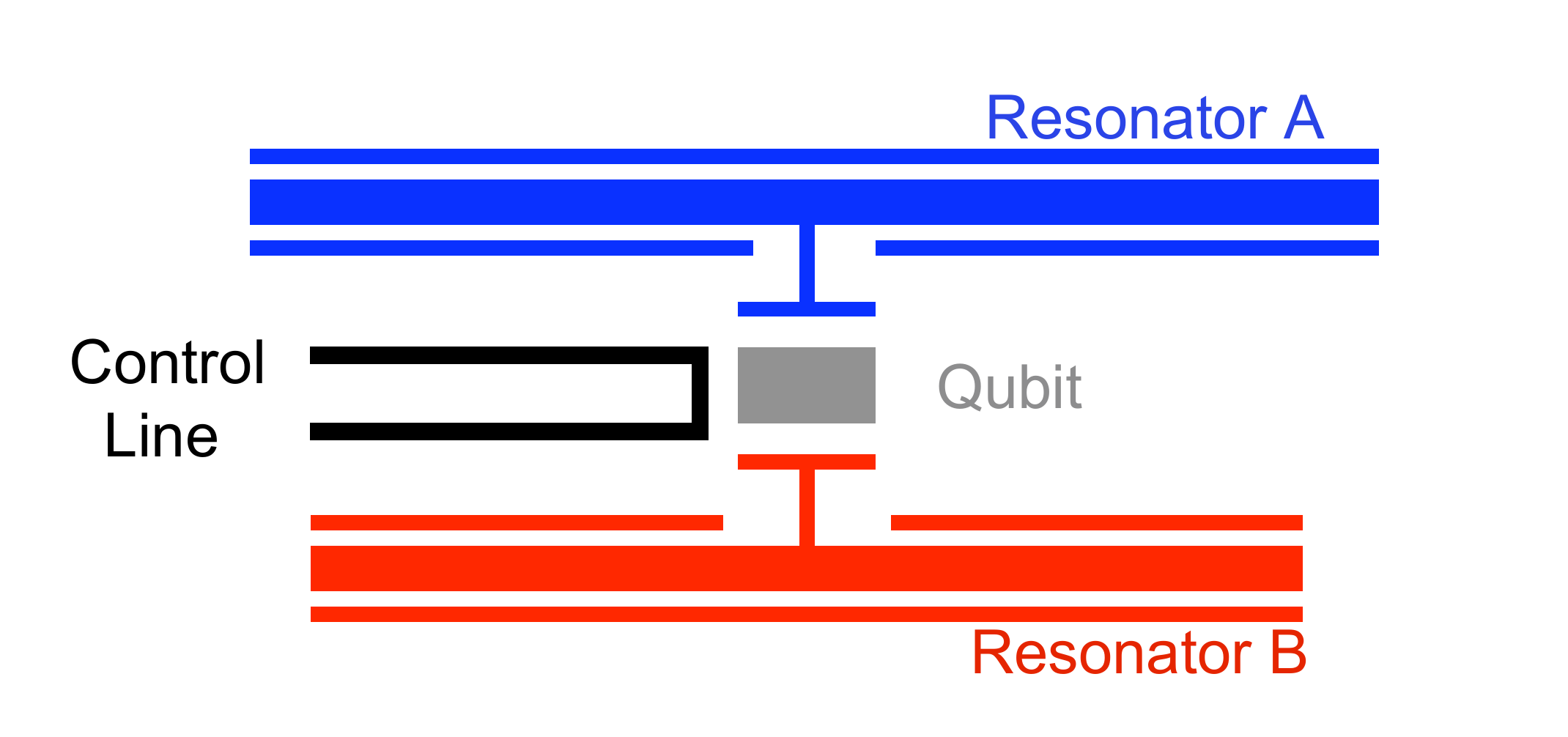}
\caption{Schematic circuit for generating entanglement between two superconducting resonators.  Resonator A (blue) has a fundamental frequency $\omega_a/2\pi$, while resonator B (red) has frequency $\omega_b/2\pi$.  These are each capacitively coupled to a tunable qubit (gray) with frequency $\omega_q/2\pi$, with coupling strengths $g_a$ and $g_b$.  The qubit is controlled by an external circuit (black).  The theoretical results described in the text require $\omega_a < \omega_q < \omega_b$.}
\label{circuitfig}
\end{center}
\end{figure}

\begin{figure}[t]
\begin{center}
\includegraphics[width=3in]{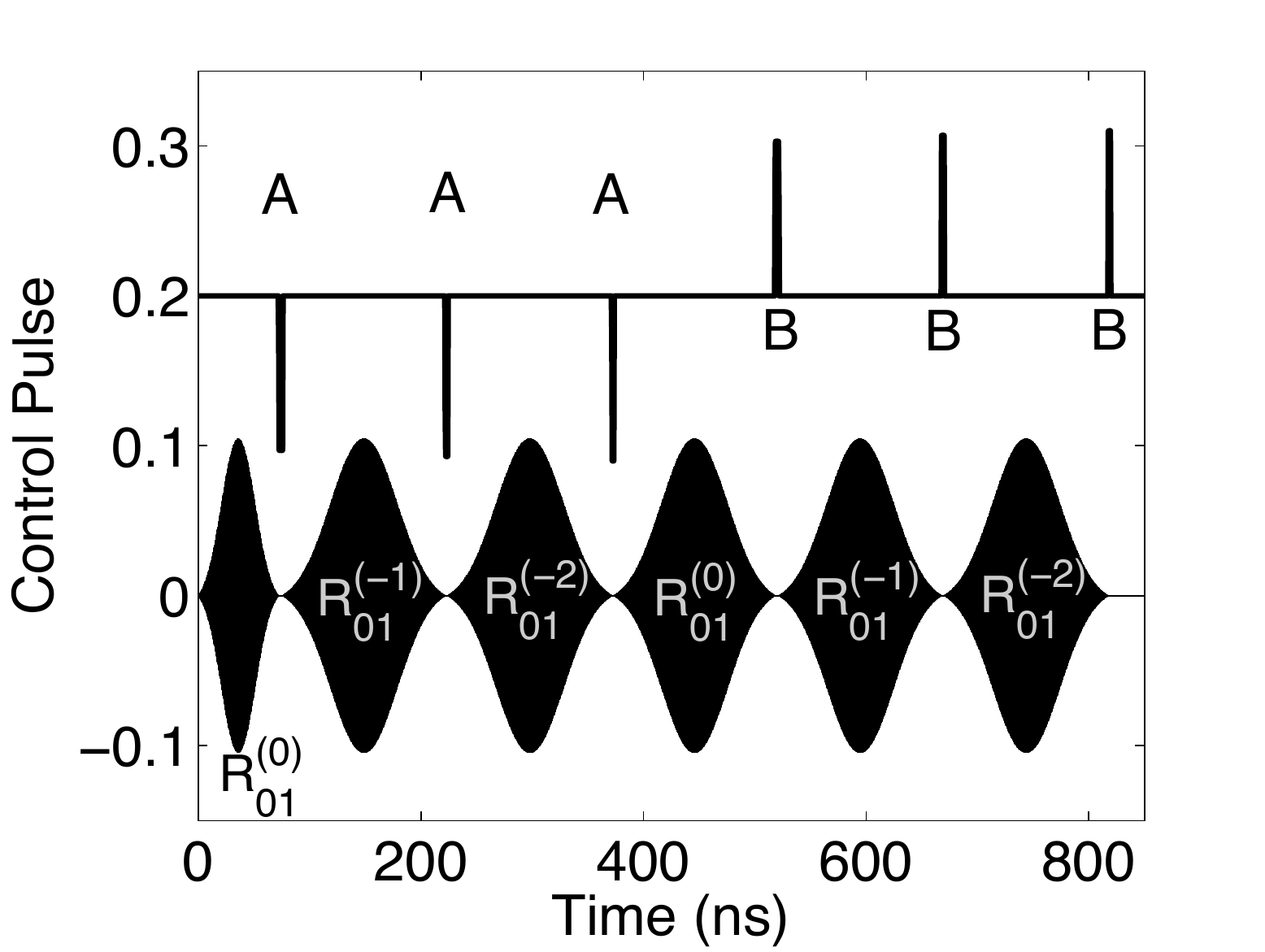}
\caption{Control Sequence for NOON State Synthesis with N =3.  The upper curve indicates shifts of the qubit frequency as a function of time, while the lower indicates the microwave pulses applied to the qubit.  Here we have set $\omega_q / 2\pi = 7 \mbox{GHz}$, $\omega_a/2\pi = 6.3 \mbox{GHz}$, $\omega_b/2\pi = 7.7 \mbox{GHz}$, and $g/2\pi = 70 \mbox{MHz}$. The various steps of the sequence are labelled, with the first Rabi transition a $\pi/2$-pulse, and the remaining all $\pi$-pulses.  }
\label{numerical1}
\end{center}
\end{figure}

The algorithm described above can be implemented by a tunable superconducting qubit, such as the phase \cite{Martinis2002} or transmon \cite{Koch07} qubit, coupled to two on-chip resonators, as shown in Fig. \ref{circuitfig}.   However, there are a number of details that will depend on the specific experimental implementation.   First, the sequence of operations must be ``programmed'' with a time-dependent control pulse for the frequency $\omega_q(t)$ and the microwave drive $\Omega(t)$.  Such pulses are shown in Fig. \ref{numerical1} for the NOON state sequence with $N=3$.

First, when analyzing the fidelity of this control sequence, it is important to note that, for fixed couplings $g_a,g_b$, the natural basis to describe the various transitions is in fact the dressed basis, i.e. those states that are true eigenstates of the Hamiltonian for $\Omega=0$ and $\omega_q$ equal to some fixed value.  While these states are close to the product states $|q,n_a,n_b\rangle$, this is only an approximation.  It is convenient to imagine that the couplings $g_a,g_b$ can be turned on and off at the beginning and end of the algorithm.  This could be achieved by using a tunable coupler \cite{Allman2010,Bialczak11} or by moving the qubit frequency to the regime $\omega_q \gg \omega_a,\omega_b$.

\begin{figure}[t]
\begin{center}
\includegraphics[width=3in]{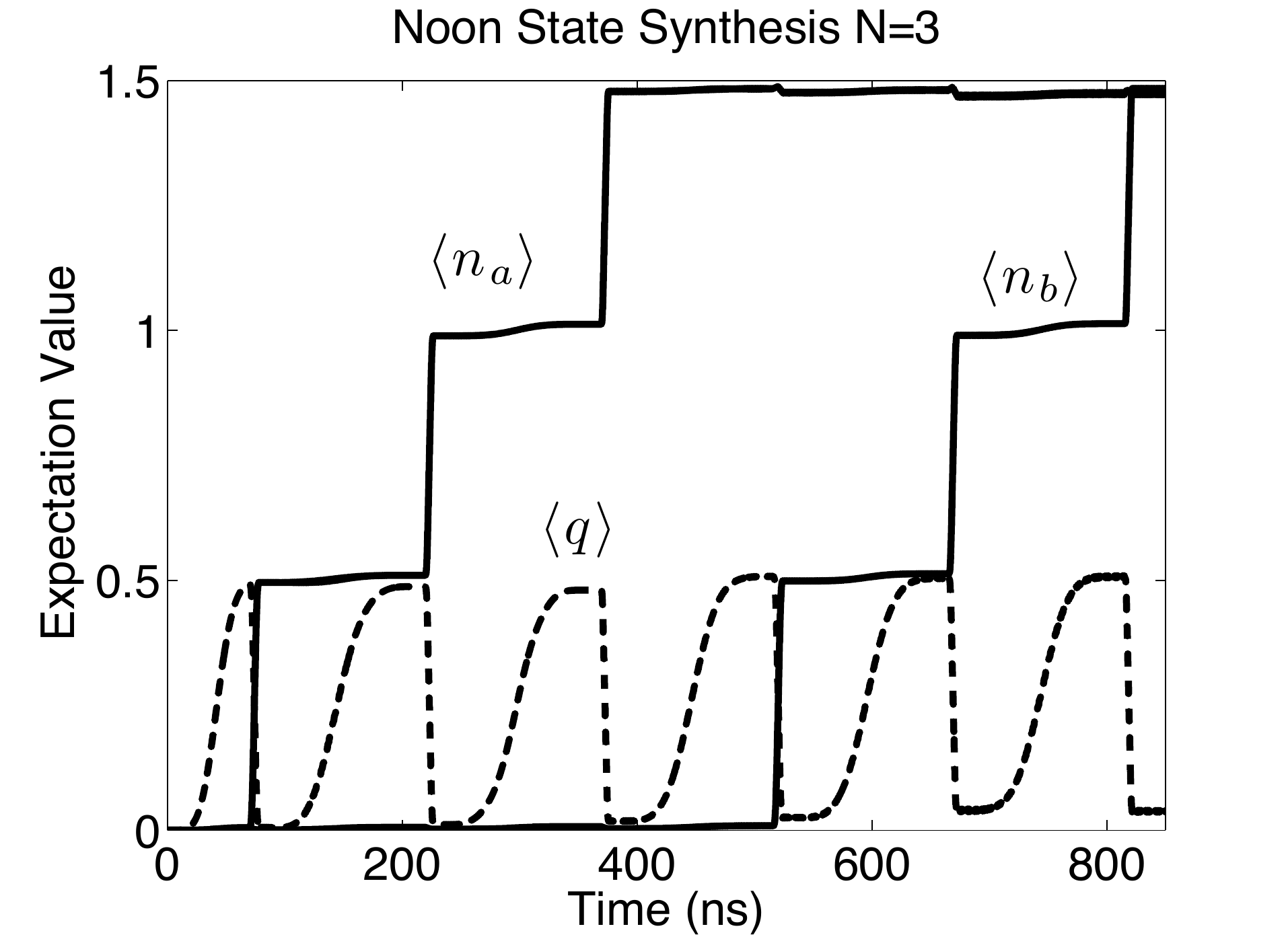}
\caption{Expectation values of the qubit excitation $\langle q\rangle$ and photon numbers $\langle n_a\rangle$ and $\langle n_b\rangle$ as a function of time for the NOON state sequence with $N=3$, using the control sequence and parameters of Fig. \ref{numerical1}.  These values have been averaged over a window of $5 \mbox{ns}$ to remove oscillations due to the dressed-state components of the wavefunction (see text).  }
\label{numerical2}
\end{center}
\end{figure}

Second, each of the number-state-dependent Rabi transitions will in fact require optimization in both amplitude and frequency for each desired Fock state $(n_a,n_b)$.   The amplitude optimization is necessary because of the variation of the matrix elements (of $\sigma_x$) between the dressed states, while the frequency optimization is necessary because the expression for $\omega_{d}$ in Eq. (\ref{drivefreq}) is only a leading-order perturbative result for a two-level system.  The results of such a partial optimization for the control pulse of Fig. \ref{numerical1} is shown in Fig. \ref{numerical2}.  Here we have shown the expectation values $\langle n_a \rangle$, $\langle n_b \rangle$, and $\langle q \rangle$ (in the uncoupled basis), averaged over a window of $5 \mbox{ns}$ to remove high-frequency oscillations (due to fixed coupling and the dressed basis).  Each Rabi pulse increases the qubit population, while each swap ($A$ or $B$) transfers the qubit population to the oscillator (thereby increasing $\langle n_a\rangle$ or $\langle n_b\rangle$).   This control sequence achieves a state equivalent to the $N=3$ NOON state with a fidelity of $0.975$.  Higher fidelity should be possible by using more sophisticated Rabi and shift pulses, or optimal control techniques.

Third, the dispersive shifts underlying the number-state-dependent transitions for a three-level system are in fact quite different than the two-level results.  For a three-level system with level spacings $\omega_{01}, \omega_{12}$, the frequency required for $|0,n_a,n_b\rangle \to |1,n_a,n_b\rangle$ becomes 
\begin{eqnarray}
\omega_{01}^{(n_a,n_b)} &\approx& \omega_{01} + \frac{g_a^2}{\omega_{01} - \omega_a} (2n_a+1) + \frac{g_a^2 \lambda^2}{\omega_a - \omega_{12}} n_a \nonumber \\
& & + \frac{g_b^2}{\omega_1 - \omega_b} (2 n_b+1) + \frac{g_b^2 \lambda^2}{\omega_b-\omega_{12}} n_b,
\end{eqnarray}
with $\lambda \approx \sqrt{2}$.  This dispersive shift depends on the relationship of both $\omega_{01}$ and $\omega_{12}$ with $\omega_a$ and $\omega_b$, with somewhat surprising results \cite{Koch07,Strauch11}.  However, many of the results described above can be adapted to handle these complications.  By setting the frequencies to $\omega_a < \omega_{12} < \omega_{01} < \omega_b$, it is possible to tune the qubit to feel an equal (not opposite) Stark shift from the two resonators, so that the resonant condition can be written as
\begin{equation}
\omega_n = \omega_q + \Delta \omega_0 + n \Delta \omega
\end{equation}
with $n = n_a + n_b$ and $\Delta \omega < 0$.  In order to properly solve the inverse equations, the only change needed is to replace the order of the product in $U_{b,j}^{\dagger}$:
\begin{equation}
U_{b,j}^{\dagger} = R_{b,jN_a}^{\dagger} B_{jNa}^{\dagger}  R_{b,j(Na-1)}^{\dagger} B_{j(N_a-1)}^{\dagger} \cdots R_{j 0}^{\dagger} B_{j0}^{\dagger}.
\end{equation}
The transitions for row $j$ are now marching along columns $k$ from left-to-right, preventing transitions to propagate back up from row $j-1$ (note that this does not affect the NOON state synthesis procedure).  Otherwise, the algorithm performs quite similarly to that described above.

%%%%%%%%%%%%%%%%%%%%%%%%%%%%%%%%%%%%%%%%%
%%%%%%%%%%%%%%%%%%%%%%%%%%%%%%%%%%%%%%%%%

%%%%%%%%%%%%%%%%%%%%%%%%%%%%%%%%%%%%%%%%
%%%%%%%%%%%%%%%%%%%%%%%%%%%%%%%%%%%%%%%%%

%%%%%%%%%%%%%%%%%%%%%%%%%%%%%%%%%%%%%%%%
%%%%%%%%%%%%%%%%%%%%%%%%%%%%%%%%%%%%%%%%%

\section{NOON State Synthesis with Decoherence}

There are two known procedures to generate the entangled state between two resonators 
\begin{equation}
|\Psi_{\subs{target}} \rangle = \frac{1}{\sqrt{2}} \left( |N,0\rangle + |0,N\rangle \right),
\end{equation}
which we shall call Method 1 (for \cite{Strauch10}) and Method 2 (for \cite{Merkel10,Wang11}).  The behavior of these two procedures under decoherence is the subject of this section.

We consider the role of dissipation on this process, modeled by a Lindbald equation
\begin{equation}
\frac{ d \rho}{d t}  = - \frac{i}{\hbar} [\mathcal{H}, \rho] + \sum_{j=1}^3 \lambda_j (L_j \rho L_j^{\dagger} - \frac{1}{2} L_j^{\dagger} L_j \rho - \frac{1}{2} \rho L_j^{\dagger} L_j)
\end{equation}
where $\lambda_1 = 1/T_q$, $\lambda_2 = \lambda_3 = 1/T_r$, $L_1 = \sigma_-$, $L_2 = a$, and $L_3 = b$.  This equation is perturbatively solved in the Appendix, where the final state of the two resonators has the approximate form
\begin{eqnarray}
\rho  &=& \rho_{a,a} |N,0\rangle \langle N,0| + \rho_{a,b} |N,0\rangle \langle 0,N\rangle \nonumber \\
& &+ \rho_{b,a} |0,N\rangle \langle N,0| + \rho_{b,b} |0,N \rangle \langle 0,N|,
\end{eqnarray}
and explicit expressions for the populations $\rho_{a,a}, \rho_{b,b}$ and the coherences $\rho_{a,b}, \rho_{b,a}$ are derived.  From these, we calculate the NOON state fidelity by
\begin{equation}
\mathcal{F} = \langle \Psi_{\subs{target}} | \rho | \Psi_{\subs{target}} \rangle = \frac{1}{2} \left( \rho_{aa} + \rho_{ab} + \rho_{ba} + \rho_{bb} \right).
\end{equation}

\begin{figure}[t]
\begin{center}
\includegraphics[width=3in]{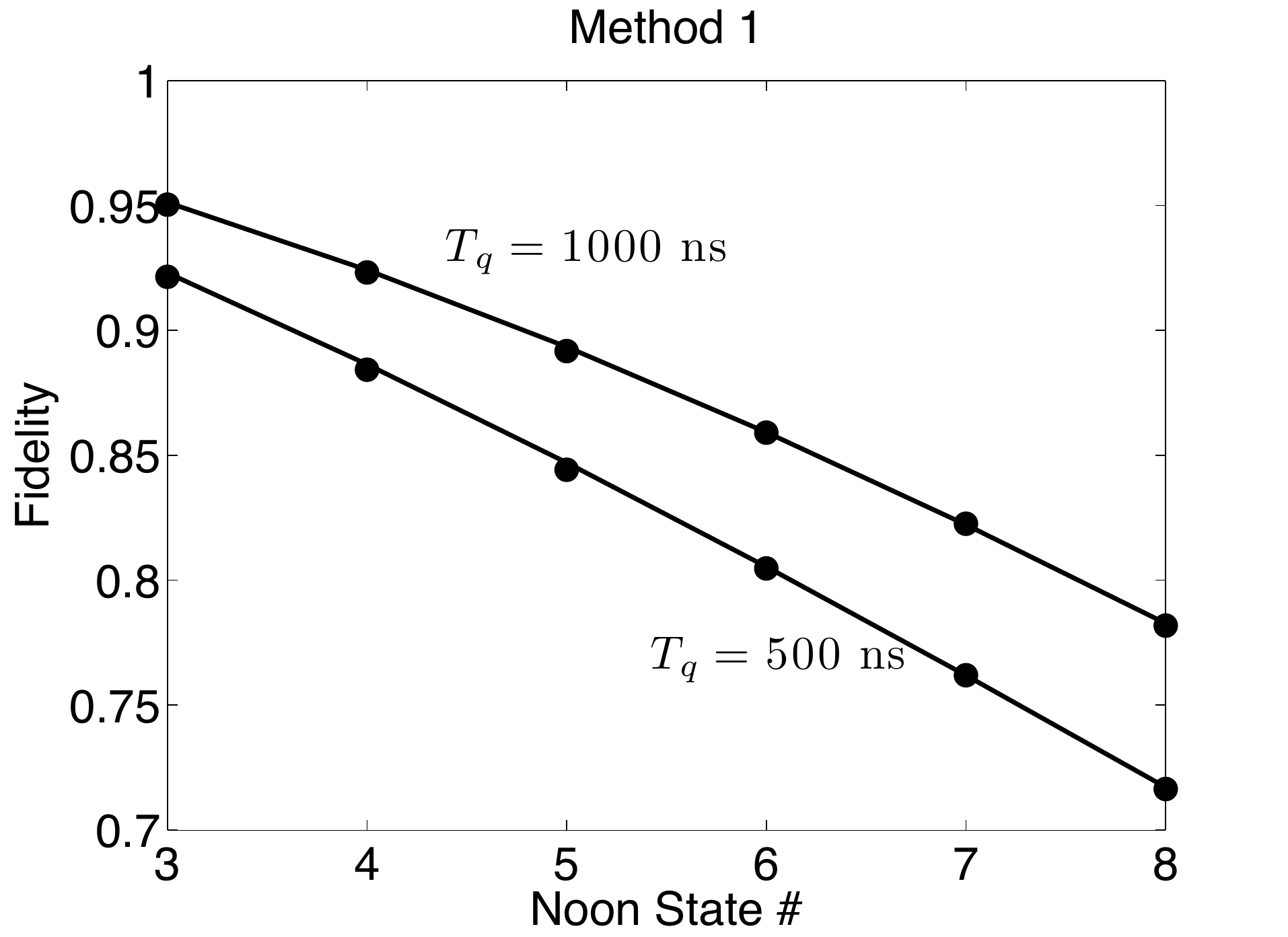}
\caption{State fidelity $\mathcal{F}_{M_1}$ as a function of $N$ for the NOON State Sequence Method 1.  The upper curve is for a qubit dissipation time $T_q = 1000 \mbox{ns}$, while the lower curve is for $T_q = 500 \mbox{ns}$.  Other parameters are resonator decoherence $T_r = 10 \mu\mbox{s}$, $\Omega/2\pi = 20 \mbox{MHz}$, and $g/2\pi = 100 \mbox{MHz}$. }
\label{decoherence1}
\end{center}
\end{figure}

Method 1 uses the state synthesis algorithm discussed above, for which the preparation sequence can be written as
$U = (B R_{01})^N (A R_{01})^N$.  For simplicity, we include only resonant interactions for each time-step, as described in \cite{Strauch11}, for both our numerical and analytical calculations.  As shown in the Appendix, the fidelity can be approximated by
\begin{eqnarray}
\mathcal{F}_{M_1} \approx \exp \left[ - \frac{7}{16} \left(N - \frac{1}{2}\right) \frac{\Delta t}{T_{q}} - N \left(N-\frac{1}{2}\right) \frac{\Delta t}{T_{r}}  \right] \nonumber \\
\times \exp \left[- \frac{1}{2} \sum_{n=1}^N \Delta t_n  \left(  \frac{1}{T_q} + \frac{2n+N-1}{T_r}\right)\right],
\end{eqnarray}
where $\Delta t = \pi/ \Omega$ and $\Delta t_n = \pi / (2 g \sqrt{n})$.  This is shown as a function of $N$ in Fig. \ref{decoherence1} for typical physical parameters, along with a numerical simulation of the Lindblad equation using a quantum trajectories method.   Each data point was calculated using 1024 trajectories with the same physical parameters.  The figure demonstrates that the perturbative result derived in the Appendix is quite good. 

While the dominant loss of fidelity in this method is due to dissipation in the qubit, there is still a significant contribution related to dissipation int he resonator.  The quadratic dependence on $N$ of the latter is due to the fact that the Fock state $|n\rangle$ decays with a rate $n/T_r$.  To a good approximation, the overall process has a fidelity of the form 
\begin{equation} 
\mathcal{F}_{M_1} \sim e^{-(7/32) T_{M_1}/T_q} e^{- (1/2) N T_{M_1}/T_r}
\end{equation}
where $T_{M_1} = 2 N \Delta t + 2 \sum_{n} \Delta t_n$ is the total duration of the sequence for Method 1. 

\begin{figure}[t]
\begin{center}
\includegraphics[width=3in]{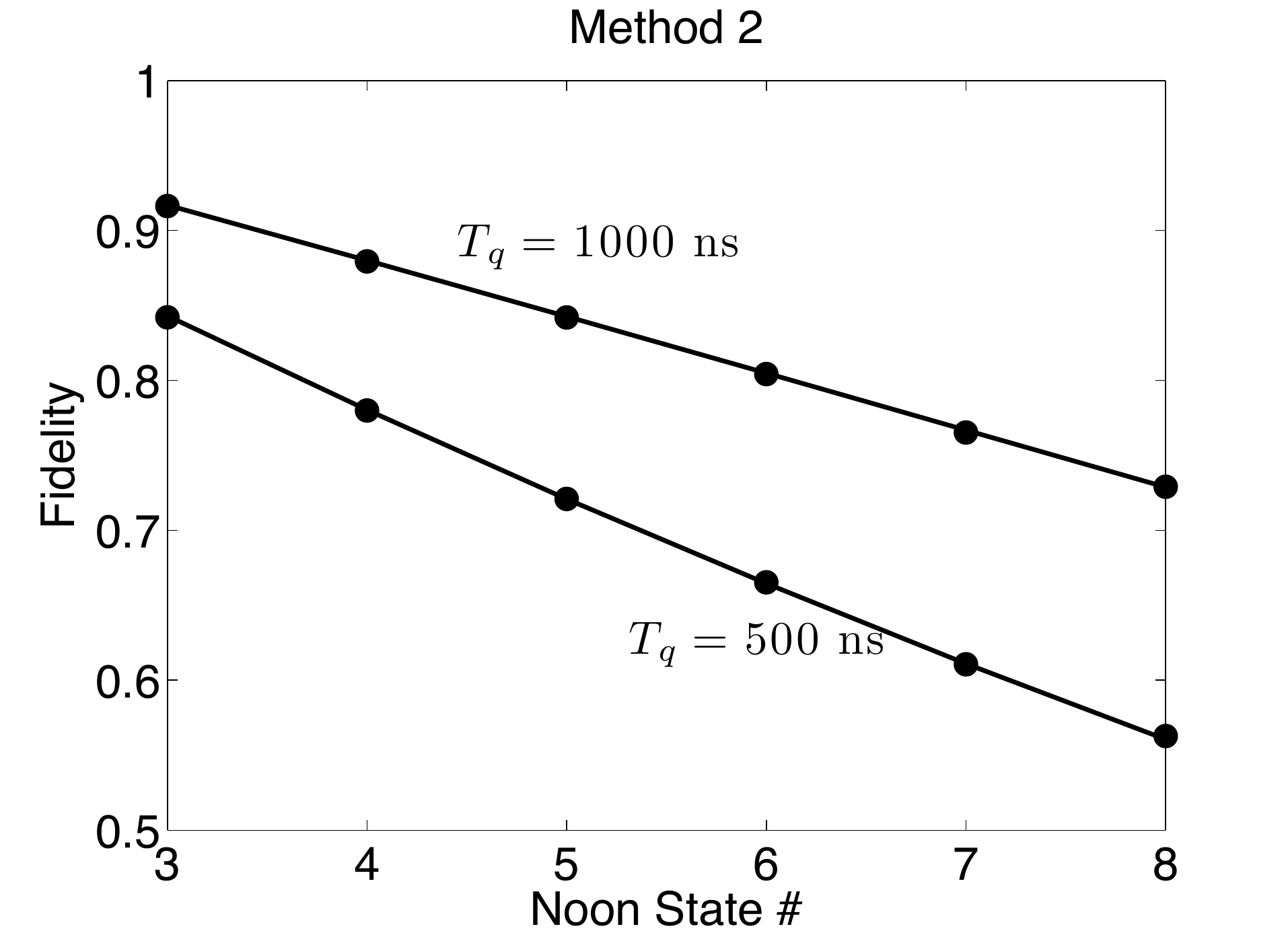}
\caption{State fidelity $\mathcal{F}_{M_2}$ as a function of $N$ for the NOON State Sequence Method 2.  The upper curve is for a qubit dissipation time $T_q = 1000 \mbox{ns}$, while the lower curve is for $T_q = 500 \mbox{ns}$.  Other parameters are resonator decoherence $T_r = 10 \mu\mbox{s}$, $\Omega/2\pi = 20 \mbox{MHz}$, and $g/2\pi = 100 \mbox{MHz}$. }
\label{decoherence2}
\end{center}
\end{figure}

The recent NOON state experiment \cite{Wang11} used a different procedure (Method 2), with two three level systems initially in the entangled state
\begin{equation}
\frac{1}{\sqrt{2}} \left( |01\rangle + |10\rangle \right).
\end{equation}
and subject to the sequence of pulses of the form $(A B) [A B R_{12}^{(a)} R_{12}^{(b)}]^{N-1}$.  Here $A$, $B$, and $R_{12}$ are detailed in the Appendix, where the fidelity for this method is shown to be approximately given by
\begin{eqnarray}
\mathcal{F}_{M_2} &\approx& \exp \left[ - \frac{11}{8} \left(N - \frac{8}{11}\right) \frac{\Delta t}{T_q} - \frac{1}{2} (N-1) (N-2)  \frac{\Delta t}{T_{r}} \right] \nonumber \\
& & \times \exp \left[ - \frac{1}{2} \sum_{n=1}^{N-1} \Delta t_n \left(\frac{3}{T_q} + \frac{2n-1}{T_r} \right)  \right] \nonumber \\
& & \times \exp \left[ - \frac{1}{2} \frac{\Delta t_0}{T_q} - \frac{1}{2} \Delta t_N \left(\frac{1}{T_q} + \frac{2N-1}{T_r}\right) \right],
\end{eqnarray}
with $\Delta t = \pi/ \Omega$, $\Delta t_n = \pi / (2 \sqrt{2 n} g)$ for $n = 1 \to N-1$, $\Delta t_0 = \pi/(4 g)$, and $\Delta t_N = \pi/(2 \sqrt{N} g)$ (see the Appendix for more details).  This fidelity is shown as a function of $N$ in Fig. \ref{decoherence2}, using the same parameters as Fig. \ref{decoherence1}.  Here the fidelity has the approximate form
\begin{equation}
 \mathcal{F}_{M_2} \sim e^{-(11/8) T_{M_2}/T_q} e^{- (1/2) N T_{M_2}/T_r}, 
 \end{equation}
 where $T_{M2} = N \Delta t + \sum_{n} \Delta t_n$ is the total duration of the sequence for Method 2.  Note that this method has a stronger dependence on the qubit coherence, due to the fact that the second-excited state decays at a rate $2/T_q$.  The total time of Method 2, however, is approximately half that of Method 1, as the $(A R)^{N-1}$ and $(B R)^{N-1}$ operations are performed in parallel.

\begin{figure}[t]
\begin{center}
\includegraphics[width=3in]{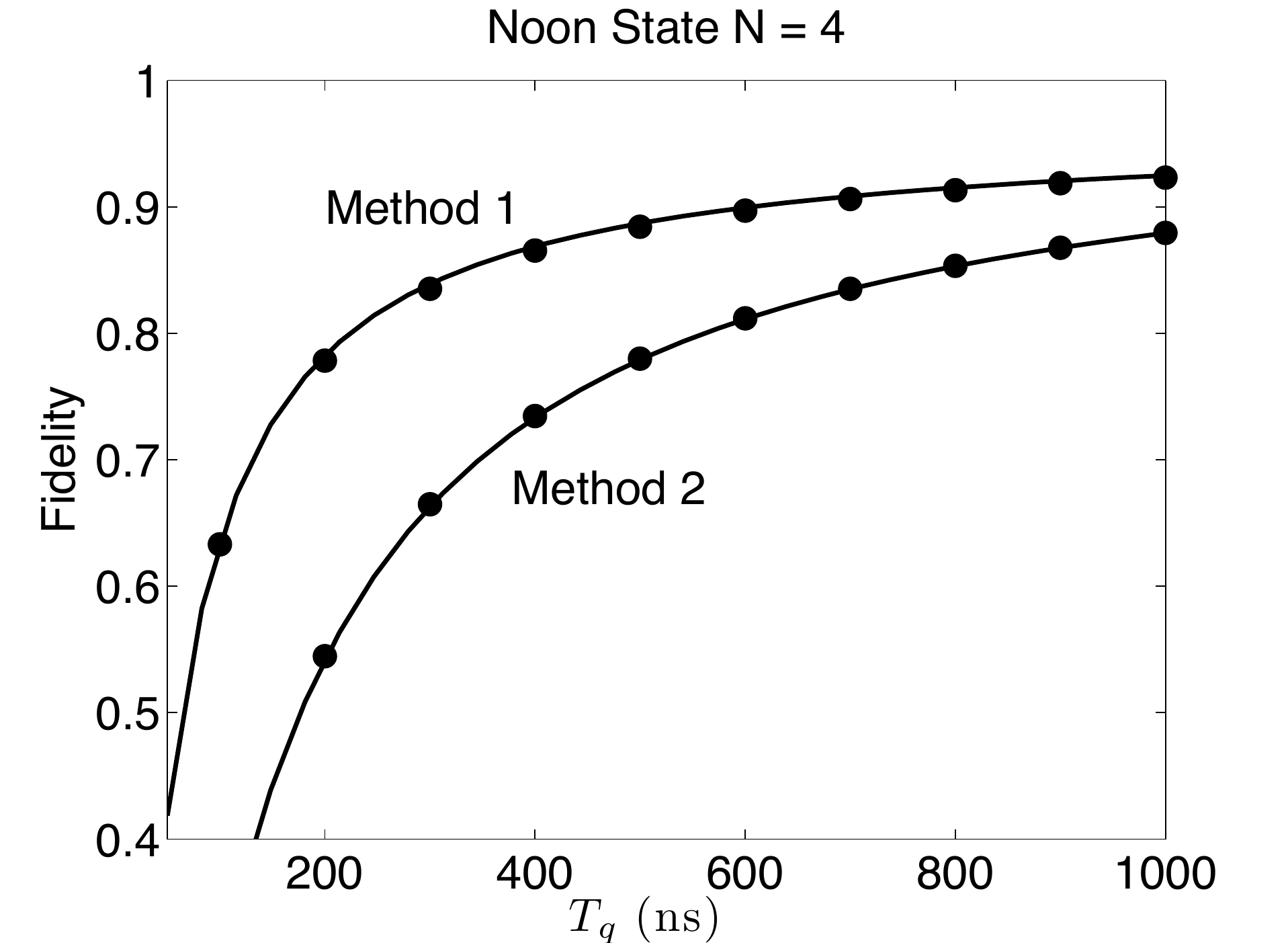}
\caption{State fidelity as a function of qubit dissipation time $T_q$ for the NOON State Methods with $N=4$.  The upper curve is for Method 1, while the lower curve is for Method 2.  Other parameters are resonator decoherence $T_r = 10000 \mbox{ns}$, $\Omega/2\pi = 20 \mbox{MHz}$, and $g/2\pi = 100 \mbox{MHz}$. }
\label{decoherence3}
\end{center}
\end{figure}

The two procedures, for $N=4$, are directly compared in Fig. \ref{decoherence3} as a function of the qubit decay time $T_q$.  For the same values of $\Delta t = \pi / \Omega$, and equal couplings $g > \Omega$,  Method 1 outperforms Method 2 for all values of $N$ and $T_q$.  While this is somewhat surprising, it should be pointed out that the second method does not use any dispersive interactions, so the Rabi pulses can be driven faster to decrease the overall preparation time even further.  

%%%%%%%%%%%%%%%%%%%%%%%%%%%%%%%%%%%%%%%%
%%%%%%%%%%%%%%%%%%%%%%%%%%%%%%%%%%%%%%%%%

%%%%%%%%%%%%%%%%%%%%%%%%%%%%%%%%%%%%%%%%
%%%%%%%%%%%%%%%%%%%%%%%%%%%%%%%%%%%%%%%%%

%%%%%%%%%%%%%%%%%%%%%%%%%%%%%%%%%%%%%%%%
%%%%%%%%%%%%%%%%%%%%%%%%%%%%%%%%%%%%%%%%%

\section{Conclusion}
In this paper we have explored the synthesis of arbitrary entangled states between two superconducting resonators.  Elsewhere, one of us has proposed quantum logic operations for such resonators as {\it qudits} \cite{Strauch11}.  Such methods, using the larger Hilbert space afforded by harmonic oscillator modes, may prove to be an important alternative to qubit operations.  The fabrication of on-chip stripline or coplanar waveguide resonators is often much easier than qubits based on Josephson junctions, and can be expected to yield coherence times of $10 \mu \mbox{s}$, much higher than the qubits.  Recent experiments \cite{Paik2011} have shown that three-dimensional cavity resonances can have even higher coherence times $T_r \approx 50 \mu\mbox{s}$.  Understanding how to use the larger resources afforded by these devices remains a primary goal.  We conclude with a discussion of important topics for future study of entangled resonators.

It is interesting to compare and contrast the results of the state-synthesis algorithm of \cite{Strauch10} (Method 1) with the NOON state preparation procedure of \cite{Merkel10,Wang11} (Method 2).  Both involve a set of Rabi and swap pulses, and exhibit similar results for the fidelity.  Method 1, however, requires the use of the number-state-dependent Rabi transitions in the dispersive regime.  These selective transitions must have Rabi frequencies less than the separation in frequencies to neighboring transitions, and thus high-fidelity operations will typically be slower than non-selective Rabi transitions used in method 2.  While we have not attempted a full optimization of either method, it is clear than a faster state synthesis algorithm is desirable, especially in the presence of decoherence.  

Entangled resonators can be used to test higher-dimensional Bell inequalities \cite{Collins2002}.  While the Bell inequality has been tested using two superconducting qubits \cite{Ansmann2009,DiCarlo2009}, and the Mermin inequality \cite{Mermin90} with three superconducting qubits \cite{DiCarlo2010, Neeley2010}, there remains much to be explored about nonlocality for coupled qudits \cite{He2011}.  HIgher-dimensional inequalities allow for both more sensitive and more robust demonstrations of quantum mechanics, and have recently been studied experimentally with optical photons \cite{Dada2011} for $d$ up to 12.  Aside from the entangled states themselves, these tests would require  qudit logic gates \cite{Strauch11} and measurements.  While qudit measurements can be accomplished using coupled qubits in either the resonant \cite{Hofheinz2009} or quasi-dispersive \cite{Johnson10} regimes, these qubit-based schemes encode the resonator state in a sequence of two-outcome results.   It would be advantageous to develop $d$-outcome measurements (with $d>2$) to more efficiently read out the resonator states.  

Entangled resonators may also be useful for quantum communication.  As resonators have the capacity to store greater entanglement than qubits, these would fit in nicely with the quantum routing protocols of \cite{Chudzicki10}.  Finally, it may be possible to extend the quantum optics analogy with circuit-QED further, and use these entangled states as interferometric probes to achieve Heisenberg-limited measurements \cite{Lee02,Dowling08} of microwave fields.   These and other probing questions will undoubtedly lead to new and improved methods for quantum control, measurement, and sensing with superconducting devices.

\acknowledgements
We gratefully acknowlege discussions with J. Aumentado, F. Altomare, and B. Johnson.  FWS was supported by the Research Corporation for Science Advancement, KJ by the NSF under Project No. PHY-0902906, and both by the NSF under Project No. PHY-1005571.  

\appendix*
\section{Decoherence Calculations}
Each step of the two NOON state synthesis procedures involves a two-state oscillation, where each state is subject to decoherence, taken to be dissipation of the qubit and oscillator with decay times $T_q$ and $T_r$, respectively (dephasing can be included similarly).  This is modeled by a reduced master equation for the density matrix for the two states $|1\rangle$ and $|2\rangle$
\begin{equation}
\rho = \rho_{11} |1\rangle \langle 1| + \rho_{12} |1\rangle \langle 2| + \rho_{21} |2\rangle \langle 1| + \rho_{22} |2\rangle \langle 2|.
\end{equation}
The master equation takes the form
\begin{eqnarray}
\frac{d \rho_{11}}{dt} &=& i \frac{\Omega}{2} (\rho_{12} - \rho_{21}) - \lambda_1 \rho_{11} + \lambda_{12} \rho_{22}, \nonumber \\
\frac{d \rho_{12}}{dt} &=& i \frac{\Omega}{2} (\rho_{11} - \rho_{22}) - \frac{1}{2} (\lambda_{1} + \lambda_{2} ) \rho_{12}, \nonumber \\
\frac{d \rho_{21}}{dt} &=& i \frac{\Omega}{2} (\rho_{22} - \rho_{11}) - \frac{1}{2} (\lambda_{1} + \lambda_{2} ) \rho_{21}, \nonumber \\
\frac{d \rho_{22}}{dt} &=& i \frac{\Omega}{2} (\rho_{21} - \rho_{12}) - \lambda_2 \rho_{22},
\end{eqnarray}
where $\Omega$ is the rotation rate between the two states and the various decoherence rates $\lambda_1, \lambda_2, \lambda_{12}$ are summarized in Table 1.   This equation can be solved approximately using standard first-order perturbation theory (for $\lambda_1,\lambda_2,\lambda_{12} \ll \Omega$) to find
\begin{eqnarray}
\rho_{11}(t) &=& a_+(t) \rho_{11}(0)  + a_-(t) \rho_{22}(0)  \nonumber \\
& & + i b(t) \left[ \rho_{12}(0) - \rho_{21}(0) \right], \nonumber \\
\rho_{12}(t) &=& c_+(t) \rho_{12}(0) + c_-(t) \rho_{21}(0) \nonumber \\
& & + i b(t) \left[ \rho_{11}(0) - \rho_{22}(0) \right], \nonumber \\
\rho_{21}(t) &=& c_+(t) \rho_{21}(0) + c_-(t) \rho_{12}(0) \nonumber \\
& & + i b(t) \left[ \rho_{22}(0) - \rho_{11}(0) \right], \nonumber \\
\rho_{22}(t) &=& a_+(t) \rho_{22}(0) + a_-(t) \rho_{11}(0) \nonumber \\
& & + i b(t) \left[ \rho_{21}(0) - \rho_{12}(0) \right],
\end{eqnarray}
with
\begin{eqnarray}
a_{\pm}(t) &=& \frac{1}{2} e^{-(\lambda_1 + \lambda_2 - \lambda_{12}) t/2} \nonumber \\
& & \pm \frac{1}{2} e^{-(2\lambda_1 + 2 \lambda_{2} + \lambda_{12})t/4} \cos(\Omega t), \\
b(t) &=& \frac{1}{2} e^{-(2\lambda_1 + 2\lambda_{2} + \lambda_{12})t/4} \sin(\Omega t), \\
c_{\pm}(t) &=& \frac{1}{2} e^{-(\lambda_{1}+\lambda_2) t/2} \nonumber \\
& &  \pm \frac{1}{2} e^{-(2\lambda_1 + 2 \lambda_{2} + \lambda_{12})t/4} \cos(\Omega t). 
\end{eqnarray}
Coherences with other states oscillate at frequency $\Omega/2$ and decay with a rate $(\lambda_1+\lambda_2)/4$.  

\begin{table}[t] 
\caption{Parameters for Perturbative Treatment of Decoherence } % title of Table 
\centering      % used for centering table 
\begin{tabular}{l l l c c c}  % centered columns (4 columns) 
\hline\hline                        %inserts double horizontal lines 
Operation & State 1 & State 2 & $\lambda_1$ & $\lambda_2$ & $\lambda_{12}$\\ [0.5ex] % inserts table 
%heading 
\hline                    % inserts single horizontal line 
$R_{01}$ & $|0,n\rangle$ & $|1,n\rangle$ & $n/T_r$ & $1/T_q+ n/T_r$ & $1/T_q$  \\ 
$R_{12}$ & $|1,n\rangle$ & $|2,n\rangle$ & $1/T_q + n/T_r $ & $2/T_q + n/T_r$ & $2/T_q$ \\
$A_1$ & $|0,n+1\rangle$ & $|1,n\rangle$ & $(n+1)/T_r$ & $1/T_q + n/T_r$ & $0$ \\
$A_2$ & $|1,n+1\rangle$ & $|2,n\rangle$ & $1/T_q + (n+1)/T_r$ & $2/T_q + n/T_r$ & $0$
%[1ex]       % [1ex] adds vertical space 
%\hline     %inserts single line 
\end{tabular} 
\label{perturbation}  % is used to refer this table in the text 
\end{table} 

\subsection{Method 1}
This method uses a qubit coupled to the two resonators, subject to a sequence of Stark-shifted Rabi transitions $R_{01}$ and swapping $A,B$ of excitations from the qubit to resonator $a$ and $b$ of the following form
\begin{equation}
\left[B R_{01}(\pi)\right]^{N} \left[ A R_{01}(\pi) \right]^{N-1} A R_{01}(\pi/2).
\end{equation}
Each of the Rabi transitions is a Stark-shifted  transition resonant for those Fock states with a given $n=n_a \pm n_b$.  For simplicity, we assume that all rotations have an equal Rabi frequency $\Omega$, and that the qubit-resonator couplings are also equal $g_a = g_b = g$.

We first look at the $(A R)^N$ sequence, using the notation $|q,n_a\rangle$ for states of the qubit $q=0,1$ and oscillator state $n_a$.  The first rotation is a $\pi/2$-pulse, and generates the following state of the qubit-oscillator system.
\begin{equation}
\rho = \rho_{00}^{(0)} |00\rangle \langle 00| + \rho_{01}^{(0)} |00\rangle \langle 10| + \rho_{10}^{(0)} |10\rangle \langle 00| + \rho_{11}^{(0)} |10\rangle \langle 10|
\end{equation}
with $\rho_{00}^{(0)} = \rho_{01}^{(0)} = \rho_{10}^{(0)} = \rho_{11}^{(0)} = 1/2$.  The perturbative corrections are in fact negligible for this rotation, and we have ignored any overall phases of the coherence terms.  The excitation is then swapped in time $\Delta t_1 = \pi/(2g)$ from the qubit to the oscillator, leading to 
\begin{equation}
\rho = \rho_{00}^{(1)} |00\rangle \langle 00| + \rho_{01}^{(1)} |00\rangle \langle 01| + \rho_{10}^{(1)} |01\rangle \langle 00| + \rho_{11}^{(1)} |01\rangle \langle 01|
\end{equation}
with 
\begin{eqnarray}
\rho_{00}^{(1)} &=& \frac{1}{2}, \nonumber \\
\rho_{01}^{(1)} &=& \frac{1}{2} \exp \left[ -\frac{\Delta t_1}{4} \left(\frac{1}{T_{q}} + \frac{1}{T_{r}}\right) \right], \nonumber \\
\rho_{10}^{(1)} &=& \frac{1}{2} \exp \left[ -\frac{\Delta t_1}{4} \left(\frac{1}{T_{q}} + \frac{1}{T_{r}}\right) \right], \nonumber \\
\rho_{11}^{(1)} &=& \frac{1}{2}  \exp \left[-\frac{\Delta t_1}{2} \left(\frac{1}{T_{q}} + \frac{1}{T_{r}}\right) \right].
\end{eqnarray}
Note that our perturbative approach includes only the decaying matrix elements, and not any population of the lower excited states.

The subsequent rotation is a Stark-shifted $\pi$-pulse, such that $|00\rangle \to |00\rangle$ and $|01\rangle \to |11\rangle$.  The effect of decoherence on this transition is somewhat different,  leading to the following approximation for the density matrix after the swap:
\begin{equation}
\rho = \rho^{(2)}_{00} |00\rangle \langle 00| + \rho^{(2)}_{01} |00\rangle \langle 02| +\rho^{(2)}_{10} |02\rangle \langle 00| + \rho^{(2)}_{11} |02\rangle \langle 02|
\end{equation}
with
\begin{eqnarray}
\rho^{(2)}_{00} &=& \rho^{(1)}_{00},  \nonumber \\
\rho^{(2)}_{01} &=& \exp \left[-\frac{\Delta t}{2} \left( \frac{1}{T_{r}} + \frac{1}{2T_{q}}\right) \right] \nonumber \\
& & \times \exp \left[-\frac{\Delta t_2}{4} \left(\frac{1}{T_q} + \frac{3}{T_{r}}\right) \right] \rho^{(1)}_{01}, \nonumber \\
\rho^{(2)}_{10} &=& \exp \left[-\frac{\Delta t}{2} \left( \frac{1}{T_{r}} + \frac{1}{2T_{q}}\right)\right] \nonumber \\
& & \times \exp \left[-\frac{\Delta t_2}{4} \left(\frac{1}{T_q} + \frac{3}{T_{r}}\right) \right] \rho^{(1)}_{10}, \nonumber \\
\rho^{(2)}_{11} &=& \left( \frac{1}{2} +\frac{1}{2} e^{-3 \Delta t/ 4T_q} \right) e^{- \Delta t/T_r} \nonumber \\
& & \times  \exp \left[-\frac{\Delta t_2}{2} \left(\frac{1}{T_{q}} + \frac{3}{T_{r}}\right) \right] \rho^{(1)}_{11}.
\end{eqnarray}
Here $\Delta t = \pi/\Omega$ is the time for a $\pi$-pulse, the swap time $\Delta t_2 = \Delta t_1/\sqrt{2}$ has decreased, due to the matrix elements of the Jaynes-Cummings interaction, and the resonator coherence time has also decreased.  

Continuing the sequence of $\pi$-pulses we find that after $n$ steps the state is
\begin{equation}
\rho = \rho^{(n)}_{00} |00\rangle \langle 00| + \rho^{(n)}_{01} |00\rangle \langle 0n| +\rho^{(n)}_{10} |0n\rangle \langle 00| + \rho^{(n)}_{11} |0n\rangle \langle 0n|
\end{equation}
with the recursion relation
\begin{eqnarray}
\rho^{(n+1)}_{00} &=& \rho^{(n)}_{00},   \nonumber \\
\rho^{(n+1)}_{01} &=& \exp \left[-\frac{\Delta t}{2} \left( \frac{n}{T_r} + \frac{1}{2T_q}\right) \right] \nonumber \\
& & \times \exp \left[-\frac{\Delta t_{n+1}}{4} \left(\frac{1}{T_q} + \frac{2n+1}{T_r}\right) \right] \rho^{(n)}_{01}, \nonumber \\
\rho^{(n+1)}_{10} &=& \exp \left[-\frac{\Delta t}{2} \left( \frac{n}{T_r} + \frac{1}{2T_q}\right) \right] \nonumber \\
& & \times \exp \left[-\frac{\Delta t_{n+1}}{4} \left(\frac{1}{T_q} + \frac{2n+1}{T_r}\right) \right] \rho^{(n)}_{01}, \nonumber \\
\rho^{(n+1)}_{11} &=& \left( \frac{1}{2} +\frac{1}{2} e^{-3\Delta t/4T_q} \right) e^{-n \Delta t/T_r} \nonumber \\
& &  \times \exp \left[-\frac{\Delta t_{n+1}}{2} \left(\frac{1}{T_q} + \frac{2n+1}{T_r}\right) \right] \rho^{(n)}_{11}, \nonumber \\
& & 
\end{eqnarray}
and where $\Delta t_n = \Delta t_1 / \sqrt{n}$. 

Thus, after the sequence of $(AR)^N$ pulses, the final qubit-resonator state is
\begin{eqnarray}
\rho &=& \rho^{(N)}_{00} |00\rangle \langle 00| + \rho^{(N)}_{01} |00\rangle \langle 0N| \nonumber \\
& & +\rho^{(N)}_{10} |0N\rangle \langle 00| + \rho^{(N)}_{11} |0N\rangle \langle 0N|
\end{eqnarray}
with
\begin{eqnarray}
\rho^{(N)}_{00} &=& \frac{1}{2},  \nonumber \\
\rho^{(N)}_{01} &=& \frac{1}{2} \exp \left[ -\frac{1}{4} N (N-1) \frac{\Delta t}{T_r} - \frac{1}{4} \left(N-1\right) \frac{\Delta t}{T_q} \right] \nonumber \\
& & \times \exp \left[ - \frac{1}{4} \sum_{n=1}^N \Delta t_n \left( \frac{1}{T_q} + \frac{2n-1}{T_r} \right) \right], \nonumber  \\
\rho^{(N)}_{10} &=& \frac{1}{2} \exp \left[ -\frac{1}{4} N (N-1) \frac{\Delta t}{T_r} - \frac{1}{4} \left(N-1 \right) \frac{\Delta t}{T_q} \right] \nonumber \\
& & \times \exp \left[ - \frac{1}{4} \sum_{n=1}^N \Delta t_n \left( \frac{1}{T_q} + \frac{2n-1}{T_r} \right) \right], \nonumber  \\
\rho^{(N)}_{11} &=& \frac{1}{2} \left( \frac{1}{2} + \frac{1}{2} e^{-3 \Delta t/4 T_q} \right)^{N-1} \exp \left[ - \frac{1}{2} N (N-1) \frac{ \Delta t}{T_r} \right] \nonumber \\
& & \times \exp \left[ - \frac{1}{2} \sum_{n=1}^N \Delta t_n \left(\frac{1}{T_q} + \frac{2n -1}{T_r}\right) \right].
\end{eqnarray}

We now turn to the $(B R)^N$ sequence.  Extending the notation to $|q,n_a,n_b\rangle$, the initial state is
\begin{eqnarray}
\rho &=& \rho_{00}^{(N)} |000\rangle \langle 000| + \rho_{01}^{(N)} |000\rangle \langle 0N0| \nonumber \\
& & +\rho_{10}^{(N)} |0N0\rangle \langle 000| + \rho_{11}^{(N)} |0N0\rangle \langle 0N0|.
\end{eqnarray}
The $\pi$-pulse is now Stark-shifted so that the only transition is $|000\rangle \to |100\rangle$, or more generally $|00n_b \rangle \to |10n_b\rangle$.  However, the excitations of resonator $a$ continue to decay, so that the coherences get factors of $e^{-N T/2T_r}$ while the $|N\rangle \langle N|$ population gets the factor $e^{-N T/T_r}$, where  $T = N \Delta t + \sum_{n} \Delta t_n$ is the total duration of the $(B R)^N$ sequence.  Otherwise the calculation proceeds as above, with a final state
\begin{eqnarray}
\rho &=& \rho_{00} |00N\rangle \langle 00N| + \rho_{01} |00N\rangle \langle 0N0| \nonumber \\
& & +\rho_{10} |0N0\rangle \langle 00N| + \rho_{11} |0N0\rangle \langle 0N0|.
\end{eqnarray}
with density matrix elements
\begin{eqnarray}
\rho_{00} & = & \left( \frac{1}{2} + \frac{1}{2} e^{-3 \Delta t/4 T_q} \right)^N \exp \left[ - \frac{1}{2} N (N-1) \frac{\Delta t}{T_r} \right] \nonumber \\
& & \times \exp \left[ - \frac{1}{2} \sum_{n=1}^N \Delta t_n \left(\frac{1}{T_q} + \frac{2n -1}{T_r}\right) \right] \rho_{00}^{(N)}, \nonumber \\
\rho_{01} & = &  \exp \left[ - \frac{1}{4} N (N-1) \frac{\Delta t}{T_r} - \frac{1}{4} N \frac{\Delta t}{T_q} - \frac{1}{2} N \frac{T}{T_r} \right] \nonumber \\
& &\times \exp \left[ - \frac{1}{4} \sum_{n=1}^N \Delta t_n \left(\frac{1}{T_q} + \frac{2n -1}{T_r}\right) \right] \rho_{01}^{(N)}, \nonumber \\
\rho_{10} & = &  \exp \left[ - \frac{1}{4} N (N-1) \frac{\Delta t}{T_r} - \frac{1}{4} N \frac{\Delta t}{T_q} - \frac{1}{2} N \frac{T}{T_r} \right] \nonumber \\
& & \times \exp \left[ - \frac{1}{4} \sum_{n=1}^N \Delta t_n \left(\frac{1}{T_q} + \frac{2n -1}{T_r}\right) \right] \rho_{10}^{(N)}, \nonumber \\
\rho_{11} & = & \exp \left[ -N \frac{T}{T_r} \right] \rho_{11}^{(N)}.
\end{eqnarray}

Removing the qubit state, we have the desired form:
\begin{eqnarray}
\rho &=& \rho_{a,a} |N,0\rangle \langle N,0| + \rho_{a,b} |N,0\rangle \langle 0,N\rangle \nonumber \\
& & + \rho_{b,a} |0,N\rangle \langle N,0| + \rho_{b,b} |0,N\rangle \langle |0,N|.
\end{eqnarray}
with
\begin{eqnarray}
\rho_{aa} & = & \frac{1}{2} \left( \frac{1}{2} + \frac{1}{2} e^{-3 \Delta t/4 T_q} \right)^{N-1} \exp \left[ - \frac{1}{2} N \left(N-1 \right) \frac{ \Delta t}{T_r} \right] \nonumber \\
& & \times \exp \left[ - N \frac{T}{T_r} - \frac{1}{2} \sum_{n=1}^N \Delta t_n \left(\frac{1}{T_q} + \frac{2n -1}{T_r}\right) \right] ,\nonumber \\
\rho_{ab} & = & \frac{1}{2} \exp \left[ - \frac{1}{2} N \left(N-1\right) \frac{\Delta t}{T_r} - \frac{1}{2} \left(N-\frac{1}{2}\right) \frac{\Delta t}{T_q} \right] \nonumber \\
& & \times \exp \left[ - \frac{1}{2} N \frac{T}{T_r} -\frac{1}{2} \sum_{n=1}^N \Delta t_n \left(\frac{1}{T_q} + \frac{2n -1}{T_r}\right) \right], \nonumber \\
\rho_{ba} & = & \frac{1}{2} \exp \left[ - \frac{1}{2} N \left(N-1\right) \frac{\Delta t}{T_r} - \frac{1}{2} \left(N-\frac{1}{2}\right) \frac{\Delta t}{T_q}  \right] \nonumber \\
& & \times \exp \left[ - \frac{1}{2} N \frac{T}{T_r}-\frac{1}{2} \sum_{n=1}^N \Delta t_n \left(\frac{1}{T_q} + \frac{2n -1}{T_r}\right) \right],  \nonumber \\
\rho_{bb} & = & \frac{1}{2} \left( \frac{1}{2} + \frac{1}{2} e^{-3 \Delta t/4 T_q} \right)^N \exp \left[ - \frac{1}{2} N (N-1) \frac{\Delta t}{T_r} \right] \nonumber \\
& & \times \exp \left[ - \frac{1}{2} \sum_{n=1}^N \Delta t_n \left(\frac{1}{T_q} + \frac{2n -1}{T_r}\right) \right].
\end{eqnarray}

The fidelity for this state is
\begin{equation}
\mathcal{F} = \langle \Psi_{\subs{target}} | \rho | \Psi_{\subs{target}} \rangle = \frac{1}{2} \left( \rho_{aa} + \rho_{ab} + \rho_{ba} + \rho_{bb} \right).
\end{equation}
Taylor expanding the various terms and regrouping, the fidelity can be approximated by
\begin{eqnarray}
\mathcal{F} &\approx& \exp \left[ - \frac{7}{16} \left(N - \frac{1}{2}\right) \frac{\Delta t}{T_{q}} - \frac{1}{2} N \left(N-1\right) \frac{\Delta t}{T_{r}} \right] \nonumber \\
& & \times \exp \left[ - \frac{1}{2} N \frac{T}{T_r} -\frac{1}{2} \sum_{n=1}^N \Delta t_n  \left( \frac{2n-1}{T_r} + \frac{1}{T_q} \right)\right]. \nonumber \\
& & 
\end{eqnarray}
Substituting $T = N \Delta t + \sum_n \Delta t_n$ yields the result stated in the text.

\subsection{Method 2}

In this method, two three-level systems (qutrits) are initially prepared in a two-qubit Bell state
\begin{equation}
\frac{1}{\sqrt{2}} \left( |01\rangle + |10\rangle \right).
\end{equation}
Then, each qutrit is rotated from $|1\rangle \to |2\rangle$, followed by qubit-resonator swaps ($A^{(2)}$ and $B^{(2)}$) of the form $|2,n\rangle \to |1,n+1\rangle$.  This sequence can be written as
\begin{equation}
A^{(1)} B^{(1)} \left[  A^{(2)} B^{(2)} R_{12}^{(a)} R_{12}^{(b)} \right]^{N-1},
\end{equation}
where $R_{12}$ indicates a rotation between the qutrit states $|1\rangle$ and $|2\rangle$, and the final swaps ($A^{(1)}$ and $B^{(1)}$) are of the form $|1,n\rangle \to |0,n+1\rangle$.  Note that the swaps and rotations can be performed in parallel.  For simplicity, we assume that all rotations have an equal Rabi frequency $\Omega$, and that the qubit-qubit and qubit-resonator couplings are all equal to $g$. 

We first observe that the two-qutrit Bell state can be formed by a $\pi$-pulse $|00\rangle \to |10\rangle$ followed by a partial (``square-root of'') swap $|10\rangle \to |10\rangle + |01\rangle$ (again ignoring any phases in the various operations).  Including dissipation, the initial two-qutrit density matrix is approximately
\begin{equation}
\frac{p}{2} \left( |01\rangle \langle 01\rangle +  |01\rangle \langle 10| +  |10\rangle \langle 01| + |10\rangle \langle 10| \right)
\end{equation}
where
\begin{equation}
p = \frac{1}{2} \left( 1 + e^{-3\Delta t/4 T_q} \right) e^{- \Delta t_0/T_q}
\end{equation}
with $\Delta t = \pi/\Omega$ and $\Delta t_0 = \pi/(4 g)$.

Extending the notation to $|q_a, q_b, n_a, n_b\rangle$, the first $R_{12}$ pulse yields the state
\begin{eqnarray}
\rho &=& \rho_{00}^{(0)} |0200\rangle \langle0200| + \rho_{01}^{(0)} |0200\rangle \langle 2000| \nonumber \\
& & + \rho_{10}^{(0)} |2000\rangle \langle 0200| + \rho_{11}^{(0)} |2000\rangle \langle 2000|
\end{eqnarray}
with density matrix elements
\begin{eqnarray}
\rho_{00}^{(0)} &=& \frac{p}{2} \left( \frac{1}{2} e^{-\Delta t/2T_q} + \frac{1}{2} e^{-2 \Delta t/T_q} \right), \nonumber \\
\rho_{01}^{(0)} &=& \frac{p}{2} e^{-3 \Delta t/2T_q}, \nonumber \\
\rho_{10}^{(0)} &=& \frac{p}{2} e^{-3 \Delta t/2T_q}, \nonumber \\
\rho_{11}^{(0)} &=& \frac{p}{2} \left( \frac{1}{2} e^{-\Delta t/2T_q} + \frac{1}{2} e^{-2 \Delta t/T_q} \right).
\end{eqnarray}
This is then swapped to the resonators, so that the total state becomes
\begin{eqnarray}
\rho &=& \rho_{00}^{(1)} |0101\rangle \langle 0101| + \rho_{01}^{(1)} |0101\rangle \langle 1010| \nonumber \\
& & + \rho_{10}^{(1)} |1010\rangle \langle 0101| + \rho_{11}^{(1)} |1010\rangle \langle 1010|
\end{eqnarray}
with
\begin{eqnarray}
\rho^{(1)}_{00} &=&\frac{p}{2} \left( \frac{1}{2} e^{-\Delta t/2 T_{q}} + \frac{1}{2} e^{-2 \Delta t / T_{q}}\right) \nonumber \\
& & \times \exp \left[ - \frac{\Delta t_1}{2} \left( \frac{3}{T_q} + \frac{1}{T_r} \right) \right], \nonumber \\
\rho^{(1)}_{01} &=&\frac{p}{2} e^{-3 \Delta t/2T_q} \exp \left[ - \frac{\Delta t_1}{2} \left( \frac{3}{T_q} + \frac{1}{T_r} \right) \right], \nonumber \\
\rho^{(1)}_{10} &=&\frac{p}{2} e^{-3 \Delta t/2T_q} \exp \left[ - \frac{\Delta t_1}{2} \left( \frac{3}{T_q} + \frac{1}{T_r} \right) \right],  \nonumber \\
\rho^{(1)}_{11} &=&\frac{p}{2} \left( \frac{1}{2} e^{-\Delta t/2 T_{q}} + \frac{1}{2} e^{-2 \Delta t / T_{q}}\right) \nonumber \\
& & \times \exp \left[ - \frac{\Delta t_1}{2} \left( \frac{3}{T_q} + \frac{1}{T_r} \right) \right],
\end{eqnarray}
where $\Delta t_1 = \pi/(2 \sqrt{2} g)$.  The factor of $\sqrt{2}$ in $\Delta t_1$ is due to the qutrit matrix element $\langle 1|\sigma_-|2\rangle$ \cite{Strauch11}.

After $n$ steps of this sequence, the state becomes
\begin{eqnarray}
\rho &=& \rho_{00}^{(n)} |010n\rangle \langle 010n| + \rho_{01}^{(n)} |010n\rangle \langle 10n0|  \nonumber \\
& & + \rho_{10}^{(n)} |10n0\rangle \langle 010n| + \rho_{11}^{(n)} |10n0\rangle \langle 10n0|
\end{eqnarray}
where the density matrix elements satisfy the recursion relation
\begin{eqnarray}
\rho_{00}^{(n+1)} &=&  \left( \frac{1}{2} e^{-\Delta t/2 T_{q}} + \frac{1}{2} e^{-2 \Delta t / T_{q}}\right) \times e^{-n \Delta t/T_r} \nonumber \\
&  &  \times \exp \left[ - \frac{\Delta t_{n+1}}{2} \left( \frac{3}{T_q} + \frac{2n+1}{T_r} \right) \right] \rho_{00}^{(n)}, \nonumber \\
\rho_{01}^{(n+1)} &=& e^{-3 \Delta t/2T_q} e^{-n \Delta t/T_r} \nonumber \\
& &  \times \exp \left[ - \frac{\Delta t_{n+1}}{2} \left( \frac{3}{T_q} + \frac{2n+1}{T_r} \right) \right] \rho_{01}^{(n)}, \nonumber \\
\rho_{10}^{(n+1)} &=& e^{-3 \Delta t/2T_q} e^{-n \Delta t/T_r}  \nonumber \\
& & \times \exp \left[ - \frac{\Delta t_{n+1}}{2} \left( \frac{3}{T_q} + \frac{2n+1}{T_r} \right) \right] \rho_{10}^{(n)}, \nonumber \\
\rho_{11}^{(n+1)} &=&  \left( \frac{1}{2} e^{-\Delta t/2 T_{q}} + \frac{1}{2} e^{-2 \Delta t / T_{q}}\right) e^{-n \Delta t/T_r}  \nonumber \\
& & \times \exp \left[ - \frac{\Delta t_{n+1}}{2} \left( \frac{3}{T_q} + \frac{2n+1}{T_r} \right) \right] \rho_{11}^{(n)}, \nonumber \\
\end{eqnarray}
and $\Delta t_n = \pi/(2 \sqrt{2 n} g)$.  This sequence repeats until $n=N-1$.  

The final step swaps the remaining qubit excitations into the resonators, yielding the state
\begin{eqnarray}
\rho &=& \rho_{00}^{(N)} |000N\rangle \langle 000N| + \rho_{01}^{(N)} |000N\rangle \langle 00N0| \nonumber \\
& &  + \rho_{10}^{(N)} |00N0\rangle \langle 000N| + \rho_{11}^{(N)} |00N0\rangle \langle 00N0| \nonumber \\
\end{eqnarray}
with
\begin{eqnarray}
\rho_{00}^{(N)} &=& \exp \left[ - \frac{\Delta t_N}{2} \left( \frac{1}{T_q} + \frac{2N-1}{T_r} \right) \right] \rho_{00}^{(N-1)}, \nonumber \\
\rho_{01}^{(N)} &=& \exp \left[ - \frac{\Delta t_N}{2} \left( \frac{1}{T_q} + \frac{2N-1}{T_r} \right) \right] \rho_{01}^{(N-1)}, \nonumber \\
\rho_{10}^{(N)} &=& \exp \left[ - \frac{\Delta t_N}{2} \left( \frac{1}{T_q} + \frac{2N-1}{T_r} \right) \right] \rho_{10}^{(N-1)}, \nonumber \\
\rho_{11}^{(N)} &=& \exp \left[ - \frac{\Delta t_N}{2} \left( \frac{1}{T_q} + \frac{2N-1}{T_r} \right) \right] \rho_{11}^{(N-1)}, \nonumber \\
\end{eqnarray}
and $\Delta t_N = \pi/(2 \sqrt{N} g)$.  This leaves the qutrits in their ground states, and tracking the various decoherence factors we find
\begin{eqnarray}
\rho_{aa} &=& \frac{p}{2} \left( \frac{1}{2} e^{-\Delta t/2 T_{q}} + \frac{1}{2} e^{-2 \Delta t / T_{q}}\right)^{N-1} \nonumber \\
& & \times \exp \left[ - \frac{1}{2} (N-1)(N-2) \frac{\Delta t}{T_r} \right] \nonumber \\
& & \times \exp \left[ - \frac{1}{2} \sum_{n=1}^{N} \Delta t_n \left(\frac{3-2 \delta_{n,N} }{T_q} + \frac{2n-1}{T_r} \right) \right], \nonumber \\
\rho_{ab} &=& \frac{p}{2} \exp \left[ - \frac{3}{2} (N-1) \frac{\Delta t}{T_q} - \frac{1}{2} (N-1)(N-2) \frac{\Delta t}{T_r} \right] \nonumber \\
& & \times \exp \left[ - \frac{1}{2} \sum_{n=1}^{N} \Delta t_n \left(\frac{3-2 \delta_{n,N} }{T_q} + \frac{2n-1}{T_r} \right) \right], \nonumber \\
\rho_{ba} &=& \frac{p}{2} \exp \left[ - \frac{3}{2} (N-1) \frac{\Delta t}{T_q} - \frac{1}{2} (N-1)(N-2) \frac{\Delta t}{T_r} \right] \nonumber \\
& & \times \exp \left[ - \frac{1}{2} \sum_{n=1}^{N} \Delta t_n \left(\frac{3-2 \delta_{n,N} }{T_q} + \frac{2n-1}{T_r} \right) \right], \nonumber \\
\rho_{bb} &=& \frac{p}{2} \left( \frac{1}{2} e^{-\Delta t/2 T_{q}} + \frac{1}{2} e^{-2 \Delta t / T_{q}}\right)^{N-1} \nonumber \\
& & \times \exp \left[ - \frac{1}{2} (N-1)(N-2) \frac{\Delta t}{T_r} \right] \nonumber \\
& & \times \exp \left[ - \frac{1}{2} \sum_{n=1}^{N} \Delta t_n \left(\frac{3-2 \delta_{n,N} }{T_q} + \frac{2n-1}{T_r} \right) \right]. \nonumber \\
\end{eqnarray}

The fidelity for this state is
\begin{equation}
\mathcal{F} = \langle \Psi_{\subs{target}} | \rho | \Psi_{\subs{target}} \rangle = \frac{1}{2} \left( \rho_{aa} + \rho_{ab} + \rho_{ba} + \rho_{bb} \right).
\end{equation}
Taylor expanding the various terms and regrouping, the fidelity can be approximated by
\begin{eqnarray}
\mathcal{F} &\approx& \exp \left[ - \frac{11}{8} \left(N - \frac{8}{11}\right) \frac{\Delta t}{T_q} - \frac{1}{2} (N-1) (N-2)  \frac{\Delta t}{T_{r}} \right] \nonumber \\
& & \times \exp \left[ - \frac{1}{2} \sum_{n=1}^{N-1} \Delta t_n \left(\frac{3}{T_q} + \frac{2n-1}{T_r} \right)  \right] \nonumber \\
& & \times \exp \left[ - \frac{1}{2} \frac{\Delta t_0}{T_q} - \frac{1}{2} \Delta t_N \left(\frac{1}{T_q} + \frac{2N-1}{T_r}\right) \right]. \nonumber \\
\end{eqnarray}

\bibliography{report}

\end{document}